\documentclass[preprint, 
superscriptaddress,
 amsmath,amssymb,
 aps,
floatfix
]{revtex4-1}

\usepackage{graphicx}% Include figure files
\usepackage{dcolumn}% Align table columns on decimal point
\usepackage{bm}% bold math
\usepackage{hyperref}% add hypertext capabilities
\usepackage{upgreek}
\usepackage{xcolor}
\usepackage[version=3]{mhchem}

\begin{document}

\title{Thermodynamic effects of solid electrolyte interphase formation from solvation and ionic association in water-in-salt electrolytes}

\author{Daniel M. Markiewitz}
\email{dmm385@mit.edu}
\affiliation{\small	Department of Chemical Engineering, Massachusetts Institute of Technology, Cambridge, Massachusetts 02139, United States\normalsize}

\author{Michael McEldrew}
\affiliation{\small Department of Chemical Engineering, Massachusetts Institute of Technology, Cambridge, Massachusetts 02139, United States\normalsize}

\author{Conor M. E. Phelan}
\affiliation{\small	Department of Materials, University of Oxford, Parks Road, Oxford OX1 3PH, United Kingdom\normalsize}

\author{Qianlu Zheng}
\affiliation{\small	Department of Civil and Environmental Engineering, The Grainger College of Engineering, University of Illinois Urbana-Champaign, Urbana, Illinois 61801, United States\normalsize}

\author{Jasper Singh}
\affiliation{\small	Department of Chemical Engineering, Massachusetts Institute of Technology, Cambridge, Massachusetts 02139, United States\normalsize}
\affiliation{\small	Department of Materials, University of Oxford, Parks Road, Oxford OX1 3PH, United Kingdom\normalsize}

\author{Robert S. Weatherup}
\affiliation{\small	Department of Materials, University of Oxford, Parks Road, Oxford OX1 3PH, United Kingdom\normalsize}
\affiliation{\small	Diamond Light Source, Didcot, Oxfordshire OX11 0DE, United Kingdom\normalsize}
\affiliation{\small	The Faraday Institution, Quad One, Harwell Science and Innovation Campus, Didcot OX11 0RA, UK\normalsize}

\author{Rosa M. Espinosa-Marzal}
\affiliation{\small	Department of Civil and Environmental Engineering, The Grainger College of Engineering, University of Illinois Urbana-Champaign, Urbana, Illinois 61801, United States\normalsize}
\affiliation{\small	Department of Materials Science and Engineering, The Grainger College of Engineering, University of Illinois Urbana-Champaign, Urbana, Illinois 61801, United States\normalsize}

\author{Martin Z. Bazant}
\email{Bazant@mit.edu}
\affiliation{\small Department of Chemical Engineering, Massachusetts Institute of Technology, Cambridge, Massachusetts 02139, United States\normalsize}
\affiliation{\small Department of Mathematics, Massachusetts Institute of Technology, Cambridge, Massachusetts 02139, United States\normalsize}

\author{Zachary A. H. Goodwin}
\email{zac.goodwin@materials.ox.ac.uk}
\affiliation{\small Department of Materials, University of Oxford, Parks Road, Oxford OX1 3PH, United Kingdom\normalsize}
\affiliation{\small	John A. Paulson School of Engineering and Applied Sciences, Harvard University, Cambridge, Massachusetts 02138, United States\normalsize}

\date{\today}

\begin{abstract}
Water-in-salt electrolytes (WiSEs) are a promising class of next-generation electrolytes. Unlike classical dilute electrolytes or more conventional battery electrolytes, WiSEs are characterized by their superconcentrated salt and low water content, which give rise to an expanded electrochemical stability window (ESW). The expansion of the ESW is due in part to the formation of an inorganic solid electrolyte interphase (SEI) that passivates the anode; this principle also proves important in commercial Li-ion batteries where graphite and Li-metal anodes operate at potentials outside the ESW of conventional carbonate electrolytes, as well as extending beyond Li-ion technologies. Solvation structure and ionic association are key descriptors for understanding the expansion of the ESW. In particular, because the reactions that lead to SEI formation, or to cathode electrolyte interphase (CEI) formation, occur at the electrode-electrolyte interface, the distribution of reactants and their solvation environments is critical. In the absence of reactions, this interfacial distribution is referred to as the electrical double layer (EDL). Here, we further develop and analyze a recently proposed thermodynamic theory of hydration and ionic association in the EDL of WiSEs. We parameterize the theory using bulk molecular dynamics simulations and benchmark it against EDL simulations, finding good qualitative agreement. Using this thermodynamic framework, we rationalize changes in the ESW through changes in bulk electrolyte activity, via the Nernst equation, which directly alters electrolyte stability, and through thermodynamic effects on reaction kinetics, via the concentration of reactant species in the Helmholtz layer. Overall, this formalism directly links solvation and ionic aggregation to changes in reactivity, providing key insights into the thermodynamic factors that influence SEI formation in WiSEs. The framework can also be applied to other liquid electrolytes, including conventional carbonate electrolytes, solvent-in-salt systems, and Na-ion electrolytes, to understand how solvation and ionic association influence ESW expansion and SEI formation.
\end{abstract}

\maketitle

\section{Introduction}

Lithium-ion batteries (LIBs) have transformed many technologies over the past decades, including portable electronic devices, transportation, and grid-level storage of renewable energy~\cite{xu2004nonaqueous,Goodenough2013ThePerspective,xu2014electrolytes,Zheng2017Uni,li2020new,Tian2021,Kang2022Nav,Piao2022,Cheng2022Sol}. A prototypical LIB electrolyte consists of a mixture of carbonate-based solvents, such as ethylene carbonate (EC) and ethyl methyl carbonate (EMC), with LiPF$_6$ at concentrations of $\sim$1 M, often with additives to achieve the desired performance~\cite{xu2004nonaqueous,xu2014electrolytes,Xie2023}. While the cost per kWh of LIBs continues to reduce year on year, there remains a great deal of interest in the development of novel chemistries in order to improve safety, capacity, sustainability and lifetime, among other goals~\cite{xu2004nonaqueous,Goodenough2013ThePerspective,xu2014electrolytes,Zheng2017Uni,li2020new,Tian2021,Kang2022Nav,Piao2022,Cheng2022Sol}. These advances are largely being pursued by changing the electrodes of LIBs—which comprise much of the volume, weight, and cost of LIBs~\cite{Orangi2024}—to Li-metal or Si anodes~\cite{wang2022liquid,yu2020molecular,yu2022rational,Swallow2022Reveal}, higher-energy-density cathodes~\cite{li2020high}, and chemistries beyond Li-ion systems~\cite{qin2019localized,zheng2018extremely,chen2021highly}, for example. However, the electrolyte that transports Li$^+$ ions (or other active ions) between the electrodes remains fundamentally important~\cite{wang2018hybrid,zhang2018aqueous,dou2018safe,Dou2019,molinari2020chelation,lui2011salts,Suo2016,kondou2018enhanced,molinari2019transport,molinari2019general,chen202063,jiang2020high,becker2020hybrid}, as all components must function efficiently together. For instance, interphases form between the electrodes and the electrolyte: the solid electrolyte interphase (SEI) on the anode side and the cathode electrolyte interphase (CEI) on the cathode side. These interphases form through the degradation of both the electrolyte and the electrode, but their formation greatly improves Coulombic efficiency by suppressing continuous electrolyte decomposition, which is key to the long-term health of the battery~\cite{xu2004nonaqueous,Goodenough2013ThePerspective,xu2014electrolytes,Zheng2017Uni,li2020new,Tian2021,Kang2022Nav,Piao2022,Cheng2022Sol,Swallow2022Reveal}.

%Understanding SEI and CEI formation is challenging since it occurs at a solid-liquid interface, where electron transfer processes and reactions occur, over length and time scales that exceed what is feasible with conventional atomistic modeling approaches~\cite{xu2004nonaqueous,Goodenough2013ThePerspective,xu2014electrolytes,Zheng2017Uni,li2020new,Tian2021,Kang2022Nav,Piao2022,Cheng2022Sol}. Machine learning interatomic potentials (MLIPs) are promising to revolutionize this area~\cite{Ioan2022,Yao2022CRev}. Unfortunately these methods can be difficult to train for such electrolytes, often don't have long-range electrostatics and including information about reaction barriers can also be challenging~\cite{Goodwin2024ML,Yang2025ML,Zhang24ML,Juraskova2025ML}. Therefore, there is still motivation for the development of simple, analytical approaches that aid our understanding of interphase formation. 
Understanding SEI and CEI formation is challenging because it occurs at a solid-liquid interface, where electron-transfer processes and reactions take place over length and time scales that exceed what is feasible with conventional atomistic modeling approaches~\cite{xu2004nonaqueous,Goodenough2013ThePerspective,xu2014electrolytes,Zheng2017Uni,li2020new,Tian2021,Kang2022Nav,Piao2022,Cheng2022Sol}. Machine learning interatomic potentials (MLIPs) show promise for revolutionizing this area~\cite{Ioan2022,Yao2022CRev}. Unfortunately, these methods can be difficult to train for such electrolytes, often lack long-range electrostatics, and can also be challenging to incorporate information about reaction barriers~\cite{Goodwin2024ML,Yang2025ML,Zhang24ML,Juraskova2025ML}. Therefore, it is important to develop simple analytical approaches that aid our understanding of interphase formation.

It is well established in the battery community that the coordination environments of Li$^+$ ions (or other active cations) correlate with Coulombic efficiency, transport properties, and interphase formation, among other properties~\cite{xu2007solvation,von2012correlating,borodin2020uncharted,andersson2020ion,Cheng2022Sol,Wang2021corsol,Kim2021Pot,Postupna2011,Borodin2014SEI,Li2015,Skarmoutsos2015,Borodin2017Mod,Han2017MD,Ravikumar2018,Shim2018,Han2019MD,Piao2020Count,Hou2021,Wu2022}. In terms of interphase formation, the relevant reactions are often understood in terms of changes in frontier orbitals (HOMO and LUMO levels) and how these depend on the coordination structure of the electrolyte~\cite{Cresce2017,Borodin2017Mod,Beltran2020LHCE,Hou2021,Wu2022,Qisheng23JACS}. However, computing these properties for an electrolyte is costly, and even more so when interfacial environments are considered~\cite{Qisheng23JACS}. An alternative approach is to understand changes in stability and reactivity from thermodynamic effects, such as changes in activity. One example is the liquid Madelung potential proposed by Takenaka \textit{et al.}~\cite{takenaka2024liquid}, which links Li$^+$ activity to the electrostatic potential in the electrolyte. Another is the thermoreversible ionic aggregation and solvation theory of McEldrew \textit{et al.}~\cite{mceldrew2020theory}, which directly links coordination environments to activity~\cite{mceldrew2021ion}, among other properties~\cite{mceldrew2020corr,McEldrewsalt2021,Goodwin2022EDL}. These changes in activity directly translate to shifts in redox potentials through the Nernst equation, thereby thermodynamically changing the stability of the electrolyte~\cite{mceldrew2021ion}, and are linked to changes in frontier orbitals~\cite{phelan2026linking}.

Another important aspect of interphase formation is the electrical double layer (EDL), i.e., the ion distribution near the charged interface~\cite{Kornyshev2007,kilic2007a,Goodwin2017a}. This is important because the species that react to form the SEI or CEI must be present at the interface for electron-transfer reactions to occur, and reaction rates being proportional to the concentration of the reactants~\cite{Bazant2009a}. Because simulating the EDL of electrolytes is costly and challenging, there is strong motivation to develop simple theoretical approaches. Many theories for predicting EDL structure exist, ranging from simple local-density approximations~\cite{Kornyshev2007,kilic2007a,Goodwin2017a} to more sophisticated approaches~\cite{Pedro2020}. In LIBs, specific interactions between ions and solvent are important, and an approach that can capture these interactions is essential~\cite{Goodwin2022EDL}. As far as we are aware, the only theory that can consistently predict coordination environments in the EDL is the work of Goodwin, Markiewitz, and co-workers~\cite{Goodwin2022EDL,Goodwin2022Kornyshev,markiewitz2024,Markiewitz2025,GoodwinHelm2025}, which extends the earlier work of McEldrew \textit{et al.}~\cite{mceldrew2020theory,mceldrew2020corr,mceldrew2021ion,McEldrewsalt2021}. This approach has been investigated for several electrolytes, including ionic liquids (ILs)~\cite{mceldrew2020corr}, salt-in-ILs~\cite{McEldrewsalt2021,markiewitz2024}, water-in-salt electrolytes (WiSEs)~\cite{mceldrew2021ion,Markiewitz2025}, and conventional battery electrolytes~\cite{Goodwin2023,GoodwinHelm2025}.

%WiSEs are a promising class of electrolytes for applications in batteries~\cite{Suo2013,Suo2015,Wang2016,Wang2018,Yamada2016,suo2017water,dou2018safe}, owing to their improved safety from the water based solvent, but where the high 21m concentration of LiTFSI (or other IL anions) extends the electrochemical stability window (ESW) up to 4~V, in contrast to dilute water-based electrolytes~\cite{Yamada2016,vatamanu2017,mceldrew2018,borodin2020uncharted,mceldrew2021ion,sayah2022super}. Moreover, the ionic associations and hydration of Li$^+$ are known to be important in 21m, as a percolating ion networks have been found with interpenetrating water domains which still allow for facile transport of Li$^+$ ions~\cite{borodin2017liquid,zheng2018understanding,lim2018,andersson2020ion,yu2020asymmetric,zhang2020potential,mceldrew2021ion,Han2021WiSE,ichii2020solvation,li2022unconventional}. McEldrew and Markiewitz \textit{et al.}~\cite{mceldrew2018,mceldrew2021ion,Markiewitz2025} have previously investigated the bulk stability and EDL properties at lower concentrations, gaining insight into the role of ionic associations and hydration on the activity and EDL of WiSEs. However the 21~m WiSE EDL has not been investigate before with this approach~\cite{Markiewitz2025}, which is the most promising case for battery technologies.
WiSEs are a promising class of electrolytes for battery applications~\cite{Suo2013,Suo2015,Wang2016,Wang2018,Yamada2016,suo2017water,dou2018safe}. They offer improved safety because of their water-based solvent, while the high concentration of 21 m LiTFSI (or other ionic-liquid anions) extends the electrochemical stability window (ESW) to $\sim$4~V, in contrast to dilute aqueous electrolytes~\cite{Yamada2016,vatamanu2017,mceldrew2018,borodin2020uncharted,mceldrew2021ion,sayah2022super}. Moreover, ionic association and Li$^+$ hydration are known to be important in 21 m WiSEs, as percolating ionic network with interpenetrating water domains have been found, while still allowing facile Li$^+$ transport~\cite{borodin2017liquid,zheng2018understanding,lim2018,andersson2020ion,yu2020asymmetric,zhang2020potential,mceldrew2021ion,Han2021WiSE,ichii2020solvation,li2022unconventional}. McEldrew, Markiewitz, and co-workers~\cite{mceldrew2018,mceldrew2021ion,Markiewitz2025} have previously investigated bulk stability and EDL properties at lower concentrations (up to 15~m), gaining insight into the role of ionic association and hydration in determining activity and EDL structure in WiSEs. However, the EDL of 21 m WiSEs has not previously been investigated using this approach~\cite{Markiewitz2025}, despite being the most promising regime for achieving the expanded ESWs for high-energy density batteries.

%In this paper, we investigate the bulk and EDL of 21~m LiTFSI WiSE from molecular dynamics simulations (MD) and theory to gain insights into the thermodynamic effects that influence electrolyte stability and interphase formation. Initially, we outline our theory for hydration and ionic associations of LiTFSI WiSEs, and we report how we've extended our analysis to investigate the EDL properties at the 21~m concentration. This theory is validated against atomistic MD simulations of the EDL of 21~m LiTFSI, where we find good qualitative agreement. Having validated the theory further, we discuss the consequences of its thermodynamic predictions in terms of solid electrolyte interphase formation, starting from the bulk stability changes with concentration, and then moving into thermodynamic effects for the kinetics of these reactions, under certain conditions. After which, we further describe how this formalism can be generalized to other electrolytes of interest for Li-metal anodes (or otherwise), and discuss how the role of the electrode needs to be further incorporated into the model.
In this paper, we investigate the bulk and EDL properties of 21 m LiTFSI WiSEs using molecular dynamics (MD) simulations and theory, with the aim of gaining insight into the thermodynamic factors that influence electrolyte stability and interphase formation. We first outline our theory of hydration and ionic association in LiTFSI WiSEs and describe how we have extended the analysis to investigate EDL properties at 21 m concentration. We then validate the theory against atomistic MD simulations of the EDL in 21 m LiTFSI and find good qualitative agreement. Having further validated the theory, we discuss the implications of its thermodynamic predictions for solid electrolyte interphase formation, beginning with bulk stability changes as a function of concentration and then turning to thermodynamic effects on reaction kinetics under specific conditions. Finally, we describe how this formalism can be generalized to other electrolytes of interest for Li-metal anodes and related systems, and we discuss how the role of the electrode should be incorporated more fully into the model.

\section{Theory}

Here, we present an overview of the EDL theory of WiSEs with thermoreversible associations, since it is important to understand the details of the predicted changes in thermodynamic stability of the electrolyte, and its implications for SEI formation. In Refs.~\citenum{mceldrew2020theory,mceldrew2020corr,mceldrew2021ion,McEldrewsalt2021,Goodwin2022EDL,Goodwin2022Kornyshev,Goodwin2023,markiewitz2024,Markiewitz2025,GoodwinHelm2025,phelan2026linking}, the theory is outlined in more detail for WiSEs, and other electrolytes, in the bulk and at electrified interfaces, if further information is required.  

The WiSE system studied here comprises cations ($+$), anions ($-$), and water ($0$). These species are assumed to exist in an incompressible lattice-gas~\cite{mceldrew2021ion}. We define the volume of a lattice site to be that of a water molecule, $v_0$. The volumes of the other species are expressed relative to $v_0$ through the number of lattice sites they occupy, $\xi_j$ = $v_j$/$v_0$. 

To treat correlations beyond mean-field, we consider the formation of associations between cations and anions, and between cations and water~\cite{mceldrew2020theory,mceldrew2021ion}. We do not consider associations between anions and water, and between water molecules, as these have been shown to be smaller effects~\cite{mceldrew2021ion}. Cations form a maximum of $f_+$ associations, anions form a maximum of $f_-$ associations, and water forms a maximum of 1 association; referred to as the functionality of a species~\cite{mceldrew2020theory,mceldrew2021ion}. Since $f_{\pm} > 1$ for WiSEs, a polydisperse set of Cayley tree clusters form, which can be uniquely classified by the rank $lms$, i.e., the number of cations $l$, anions $m$, and water $s$ in the cluster~\cite{mceldrew2020theory,mceldrew2021ion}. The Cayley tree assumption is necessary to keep this theory analytically tractable and physically intuitive. This assumption has been shown to work well for WiSE, and a variety of other concentrated electrolytes~\cite{mceldrew2020corr,mceldrew2021ion,McEldrewsalt2021,Goodwin2023,Zhang2024,phelan2026linking}. When $f_{\pm} \geq 2$, which is typical for WiSEs, a percolating ionic network (also referred to as a gel) can form~\cite{mceldrew2020theory}. 

One of the central variables of our theory is the volume fraction of each species, $\phi_j$, where \textit{j} is $+,-,0$, or the dimensionless concentrations of species $c_j$ = $\phi_j$/$\xi_j$. A quantity we aim to find from our theory is the dimensionless concentrations of each cluster rank, $c_{lms}$. When a gel phase exists we split up the total volume fractions into sol and gel contributions, $\phi_{j} = \phi_{j}^{sol} + \phi_{j}^{gel}$, where these volume fractions are determined from Flory's post-gel convention. Finally, the dimensionless concentration of each species can be written as $c_j = \sum_{lms} i c_{lms} + c_j^{gel}$, where \textit{i} is $l,m,s$ for \textit{j} is $+,-,0$, respectively.

Following Ref.~\citenum{Markiewitz2025} the free energy functional ($\mathcal{F}$) is
\begin{align}
    \label{HEnergy} 
    \beta\mathcal{F} =& \int_V  \,d\textbf{r} \left\{-\beta \frac{\epsilon_0\epsilon_r}{2}\big(\nabla\Phi\big)^2 + \beta \rho_e \Phi  - \frac{c_{001}}{v_0} \ln\left(\frac{\sinh(\beta d |\nabla \Phi|)}{\beta d |\nabla \Phi|}\right)\right\}  \nonumber \\
    &+ \frac{1}{v_0}\int_V  \,d\textbf{r} \left\{\sum_{lms} \left(c_{lms}\ln\phi_{lms} + \beta c_{lms}\Delta_{lms}\right) + \beta\Delta^{gel}_+ c^{gel}_+ + \beta\Delta^{gel}_- c^{gel}_- + \beta\Delta^{gel}_0 c^{gel}_0\right\} \nonumber \\
    &+ \int_V  \,d\textbf{r} \left\{\Lambda \left(1-\sum_{lms}(\xi_+l+\xi_-m+s)c_{lms} - \xi_+c^{gel}_+ - \xi_-c^{gel}_- - c^{gel}_0 \right)\right\} .
\end{align}

\noindent Here $\beta = 1/k_BT$ is inverse thermal energy. The first three terms (first line of the functional) represent the electrostatic contribution to the free energy, where $\epsilon_0$ and $\epsilon_r$ are, respectively, the vacuum and relative permittivity, $\Phi$(\textbf{r}) is the electrostatic potential, $\rho_e$(\textbf{r})$ = e(c_+-c_-)/v_0$ is the charge density with $e$ being elementary charge, and $c_{001}$ and $d$ are, respectively, the dimensionless concentration of free water and its dipole moment, which is assumed to be behaving as a fluctuating Langevin dipole~\cite{mceldrew2018,Markiewitz2025}. The fourth term (first on the second line) is the ideal entropy from the clusters. The fifth term (second on the second line) is the free energy of forming clusters, where $\Delta_{lms}$ is the free energy of forming a cluster of rank $lms$. The sixth, seventh, and eighth terms (remaining terms on second line) represent the free energy of species associating with the gel, $\Delta_j^{gel}$. The final term (third line) is the Lagrange multiplier, $\Lambda$(\textbf{r}), which is used to enforce the incompressibility in the EDL~\cite{markiewitz2024,Markiewitz2025}.

We consider $\Delta_{lms}$ to have of three contributions
\begin{equation}
    \label{dfrom}
    \Delta_{lms} = \Delta^{comb}_{lms} + \Delta^{bind}_{lms} + \Delta^{conf}_{lms},
\end{equation}

\noindent where $\Delta^{comb}_{lms}$ is the combinatorial entropy, $\Delta^{bind}_{lms}$ is the binding energy, and $\Delta^{conf}_{lms}$ is the configurational entropy.

The combinatorial entropy for the WiSEs Cayley tree clusters~\cite{mceldrew2021ion,Markiewitz2025} is given by
\begin{equation}
    \label{dcomb}
    \Delta^{comb}_{lms} =- k_BT\ln\{f_+^lf_-^m W_{lms}\},
\end{equation}
\noindent where
\begin{equation}
    \label{Wlms}
    W_{lms}=\frac{(f_{+}l-l)!(f_{-}m-m)!}{l!m!s!(f_{+}l-l-m-s+1)!(f_{-}m-m-l+1)!}.
\end{equation}

The binding energy for an $lms$ cluster with $l>0$ is %simply given by,
\begin{equation}
    \label{dbind}
    \Delta^{bind}_{lms} = (l+m-1)\Delta u_{+-} + s \Delta u_{+0},
\end{equation}

\noindent where $\Delta u_{+i}$ is the energy of an association between a cation and anion or water~\cite{mceldrew2021ion,Markiewitz2025}. When $l=0$, the binding energy is zero. 

Lastly, the configurational entropy is assumed to take the form
\begin{align}
     \Delta^{conf}_{lms} 
     &= -\ln \left( \frac{\xi_+l + \xi_-m + s}{\xi_+^l\xi_-^m} \right) - (l + m -1)(T\Delta s_{+-}-1) - s(T\Delta s_{+0}-1) ,
     \label{eq:delta_config}
\end{align}

\noindent where $\Delta s_{+i}$ is the entropy of an association~\cite{mceldrew2021ion,phelan2026linking}.

We can calculate the electrochemical potential of the clusters through taking the derivative of the free energy with respect to $c_{lms}$

\begin{align}
\label{ChemPot}
    \beta \bar{\mu}_{lms} =& (l-m) \beta e \Phi - \ln\left( \frac{\sinh(\beta d |\nabla \Phi|)}{\beta d |\nabla \Phi|} \right)\delta_{l,0}\delta_{m,0}\delta_{s,1} + 1 + \ln(\bar{\phi}_{lms}) + \beta\Delta_{lms} \nonumber \\
    & - (\xi_+l+\xi_-m+s)\Lambda + (\xi_+l+\xi_-m+s)\beta\bar{g}'
\end{align}

\noindent where $\bar{g}'=\bar{c}_+^{gel}\partial\bar{\Delta}_+^{gel}+\bar{c}_-^{gel}\partial\bar{\Delta}_-^{gel}+\bar{c}_0^{gel}\partial\bar{\Delta}_0^{gel}$, with the partial derivative being with respect to $\bar{\phi}_{lms}$. An overbar indicates that the variable is the EDL version and $\Phi$ is non-zero, and the absence of it indicates bulk electrolyte, i.e., with zero $\Phi$. 

In the bulk, chemical equilibrium can be established between free species and aggregates 

\begin{equation}
\label{BulkBCEq}
    l\mu_{100} + m\mu_{010} + s\mu_{001} = \mu_{lms},
\end{equation}
\noindent which can be used to find the cluster distribution
\begin{equation}
\label{BulkClusterD}
    c_{lms}=\frac{W_{lms}}{\lambda_{+-}} \left(\psi_{100}\lambda_{+-}\right)^l \left(\psi_{010}\lambda_{+-}\right)^m \left(\phi_{001}\lambda_{+0}\right)^{s},
\end{equation}

\noindent where $\psi_{100} = f_+\phi_{100}/\xi_+$ and  $\psi_{010} = f_-\phi_{010}/\xi_-$ are dimensionless concentrations of free lattice sites, and $\lambda_{+-}$ is the cation-anion association constant and $\lambda_{+0}$ is the cation-water association constant, given by $\lambda_{+i} = \exp\{-\beta\Delta f_{+i}\}$, where $\Delta f_{+i} = \Delta u_{+i} - T\Delta s_{+i}$ is the free energy of the association. 

In order to compute the cluster distribution, we need to know the volume fractions of the \textit{free} species. These, however, are not \textit{a priori} known, and in fact, we wanted to find them from the theory instead of having them as an input. To overcome this obstacle, we introduce association probabilities, $p_{ij}$, of $i$ binding to $j$~\cite{mceldrew2020theory,mceldrew2021ion}. In an independent association approximation, we can determine the bare species volume fractions as, $\phi_{100} = \phi_+(1-p_{+-}-p_{+0})^{f_+}$, $\phi_{010} = \phi_-(1-p_{-+})^{f_-}$, and $\phi_{001} = \phi_0(1-p_{0+})$.

While we seem to have simply moved the issue, as we now need to determine the association probabilities which are also not \textit{a priori} known, we can introduce the conservation of associations and the mass action laws of the open and occupied association sites~\cite{mceldrew2020theory,mceldrew2021ion} to determine these unknowns. The conservation of cation-anion associations is
\begin{equation}
    \label{cac}
    \psi_+p_{+-} = \psi_-p_{-+} = \zeta
\end{equation}

\noindent where $\psi_+ = f_+\phi_+/\xi_+$ and $\psi_- = f_-\phi_-/\xi_-$ are, respectively, the number of cation and anion association sites per lattice site, and $\zeta$ represents the number of cation-anion associations per lattice site~\cite{mceldrew2020theory,mceldrew2021ion}. Similarly, the conservation of cation-water association is,
\begin{equation}
    \label{csc}
    \psi_+p_{+0} = \phi_0p_{0+} = \Gamma
\end{equation}

\noindent where $\Gamma$ represents the number of cation-water associations per lattice site~\cite{mceldrew2020theory,mceldrew2021ion}. The mass action law of cation-anion associations is
\begin{equation}
    \label{cal}
    \lambda_{+-}\zeta = \frac{p_{+-}p_{-+}}{(1-p_{+-}-p_{+0})(1-p_{-+})}.
\end{equation}

\noindent For cation-water associations it is
\begin{equation}
    \label{csl}
    \lambda_{+0}\Gamma = \frac{p_{+0}p_{0+}}{(1-p_{+-}-p_{+0})(1-p_{0+})}.
\end{equation}

\noindent Therefore to determine the association probabilities, all we need to know is the association constants, which is another central variable of our theory.

Since the solvation shell of Li$^+$ is often full or close to being full, i.e. the number of coordinating species is approximately $f_+$. The mass action laws become challenging to solve because $p_{+-}+p_{+0} \approx 1$ is near a singularity~\cite{mceldrew2021ion}. To overcome the singularity and facilitate a simplified solution, we can introduce the so-called ``sticky-cation'' approximation, where we assume $p_{+-}+p_{+0} = 1$~\cite{mceldrew2021ion}. The created singularities in the law of mass action laws, Eq.~\eqref{cal} \& Eq.~\eqref{csl}, can be regularized by taking the ratio of these equations,
\begin{equation}
\label{lambdaeq}
    \lambda = \frac{\lambda_{+-}}{\lambda_{+0}} = \frac{p_{-+}(1-p_{0+})}{p_{0+}(1-p_{-+})},
\end{equation}

\noindent where $\lambda$ is the cation association constant ratio. Moreover, as the sticky-cation approximation requires $s=f_+l-l-m+1$, the sticky-cation cluster distribution simplifies to

\begin{equation}
\label{sBulkClusterD}
    c_{lm}=\frac{\phi_0\alpha_0 W_{lm}}{{\lambda}}\left({\lambda} \frac{\psi_+\alpha_{+-}}{\phi_0\alpha_0}\right)^{l}\left({\lambda} \frac{\psi_-\alpha_-}{\phi_0\alpha_0}\right)^{m}
\end{equation}

\noindent where
\begin{align}
    \label{sWlm}
    W_{lm}=\frac{(f_+ l -l)!(f_- m -m)!}{l!m!(f_+l-l-m+1)!(f_-m-l-m+1)!}.
\end{align}

\noindent Here, $\alpha_0=1-p_{0+}$, $\alpha_{+-}=(1-p_{+-})^{f_+}$ and $\alpha_-=(1-p_{-+})^{f_-}$ are the fraction of free water molecules, fully hydrated cations, and free anions, respectively. 

Similarly, establishing the equilibrium between the free species and the clusters \textit{within} the EDL~\cite{Goodwin2022EDL}%, it follows
\begin{equation}
\label{EDLBCEq}
    l\bar{\mu}_{100} + m\bar{\mu}_{010} + s\bar{\mu}_{001} = \bar{\mu}_{lms},
\end{equation}
\noindent we obtain an analogous solution for the EDL cluster distribution~\cite{Goodwin2022EDL,Markiewitz2025}
\begin{equation}
\label{EDLClusterD}
    \bar{c}_{lms}=\frac{W_{lms}}{\lambda_{+-}} \left(\bar{\psi}_{100}\lambda_{+-}\right)^l \left(\bar{\psi}_{010}\lambda_{+-}\right)^m \left(\bar{\phi}_{001}\bar{\lambda}_{+0}\right)^{s},
\end{equation}
\noindent where
\begin{equation}
\label{AssoEDLp0}
    \bar{\lambda}_{+0}= \lambda_{+0}\frac{\beta d|\nabla \Phi|}{\sinh(\beta d|\nabla \Phi|)}
\end{equation}

\noindent as shown above, the cation-water association constant explicitly depends on the electric field. Since the energy of free water changes in the field, but bound water is assumed not to respond to the field~\cite{Markiewitz2025}. Analogously, we can enforce the sticky-cation approximation in the EDL with $\bar{p}_{+-}+\bar{p}_{+0} = 1$, where an equivalent set of conservation of associations and mass action law applies, and we find an analogous sticky-cation cluster distribution, $\bar{c}_{lm}$, where $\bar{\lambda}$ explicitly depends on electric field because of $\bar{\lambda}_{+0}$. 

Finally to establish an equilibrium between the bulk and the EDL, we set the electrochemical potentials of the free species equal to each other, e.g., $\bar{\mu}_{010} = \mu_{010}$~\cite{Goodwin2022EDL}. Following the work of Markiewitz and Goodwin \textit{et al.}~\cite{Goodwin2022EDL,Goodwin2022Kornyshev,GoodwinHelm2025,markiewitz2024,Markiewitz2025}, these can be expressed as so-called Boltzmann closure relations to consistently establish all chemical equilibria. For the anions, we have 
\begin{equation}
    \label{banp}
    \bar{\phi}_{010} = \phi_{010}\text{exp}(\beta e\alpha \Phi + \xi_- \Lambda),
\end{equation}

\noindent and for the free water
\begin{equation}
    \label{fwatp}
    \bar{\phi}_{001} = \phi_{001}\frac{\sinh(\beta d |\nabla \Phi|)}{\beta d |\nabla \Phi|}\text{exp}(\Lambda).
\end{equation}

\noindent Since we will use the sticky-cation approximation, we consider the equilibrium between the fully hydrated cation in the bulk and the EDL
\begin{equation}
\label{sbarec}
    \bar{\phi}_{10f_+} = \phi_{10f_+}\text{exp}(-\alpha\beta e \Phi + (\xi_+ + f_+) \Lambda)
\end{equation}

\noindent where $\phi_{10f_+} = \left(1+f_+/\xi_+\right)\phi_+(1-p_{+-})^{f_+}$. There is, however, freedom in how we choose this closure relationship. For example, we could establish the equilibrium between free cations

\begin{equation}
\bar{\phi}_{100} = \phi_{100}\text{exp}(-\alpha\beta e \Phi + \xi_+ \Lambda).
\end{equation}

\noindent While in the sticky-cation approximation, this is seemingly contradictory, since there are no free cations in the bulk or the EDL, we can regularize this problem to establish a new closure relation

\begin{equation}
    \dfrac{\bar{\phi}_{+}}{\phi_{+}} =  \left( \dfrac{\psi_-\bar{p}_{+-} [1 - p_{-+}]}{\bar{\psi}_-  p_{+-}[1 - \bar{p}_{-+}]}\right)^{f_+}\text{exp}(-\alpha\beta e \Phi + \xi_+ \Lambda).
    \label{eq:close_cat_new}
\end{equation}

Moreover in these closure relations, we introduce the potential rescaling parameter $\alpha$, introduced in Ref.~\citenum{Goodwin2017a}, to further include correlations beyond the mean-field included so far. Note that formally these closure relationships only hold in the pre-gel regime, since they are obtained from equating the pre-gel chemical potentials~\cite{Goodwin2022EDL,Markiewitz2025}. In Ref.~\citenum{Goodwin2022EDL} it was shown the gel-terms are small, and that we can extrapolate into the gel regime for ILs~\cite{Goodwin2022Kornyshev} and SiILSs~\cite{markiewitz2024,Zhang2024}. However, this extrapolation into the gel in WiSEs regime without modification has not been tested~\cite{Markiewitz2025}, and we test this assumption here.

To predict the EDL properties of WiSEs, we derive our modified Poisson-Boltzmann equation by taking the functional derivative of the free energy with respect to the electrostatic potential
\begin{equation}
    \label{mpb}
    \nabla\cdot(\epsilon\nabla \Phi) = -\rho_e = -\frac{e}{v_0}(\bar{c}_+-\bar{c}_-) ,
\end{equation}

\noindent where
\begin{equation}
    \label{epsilon}
    \epsilon = \epsilon_0\epsilon_r + d\frac{\bar{c}_{001}}{v_0}\dfrac{L(\beta d|\nabla\Phi|)}{|\nabla\Phi|}.
\end{equation}
    
\noindent Here $L(x) = \coth(x) - 1/x$ is the Langevin function. The procedure implemented to solve our system of equations coupled to the modified Poisson-Boltzmann equation is discussed in Ref.~\citenum{Markiewitz2025}. Remember, even though the gel terms are neglected in the Boltzmann closure relations, the effect of screening from the gel is still introduced through the charge density in the modified Poisson-Boltzmann equation~\cite{Goodwin2022EDL}. Some quantities will be plotted in real space relative to the Debye length, $\lambda_D$, or the inverse Debye length, $\lambda_D^{-1} = \kappa = \sqrt{e^2\beta(c_++c_-)/v_0\epsilon_0\epsilon_r}$. %The differential capacitance, $C$, can be calculated through
%\begin{equation}
%    \label{DC}
%    C = \frac{d \sigma}{d \Phi}\bigg|_{x=0} ,
%\end{equation}

%\noindent where $\sigma$ is the surface charge density, and $\Phi$ is the potential drop over the EDL, as indicated by $x=0$.

\section{Methods}

Here we further analyze the constant charge EDL molecular dynamics (MD) simulations from Refs.~\citenum{mceldrew2018} of 21~m LiTFSI WiSE. Therefore, we refer the reader to Ref.~\citenum{mceldrew2018} for all the details of those simulations. To compute associations in the MD simulations, we used a real-space cut-off of 2.7 \AA\ between Li$^+$ and O's in TFSI$^-$ and H$_2$O~\cite{mceldrew2018,Markiewitz2025}. For TFSI$^-$, we count the associations based on unique anions, so bidentate associations only count as a single association in this model. To calculate the aggregates, we construct an association adjacency matrix. In the EDL, we refer to reader to Ref.~\citenum{Markiewitz2025} for a detailed discussion of how associations are distributed in real-space.

In this work for the volume fractions, we use the same values found in Ref.~\citenum{mceldrew2021ion}, $\xi_+ = 0.4$ \& $\xi_- = 10.8$. Moreover, we pick $f_+$ = 4 and $f_-=3$~\cite{mceldrew2021ion}. Using these functionalities and the coordination numbers, we can compute the association probabilities through 
\begin{equation}
    \label{sim_prob}
    p_{ij} = \left<\frac{\# \text{ of associations of type } ij}{f_i \cdot \# \text{ of molecules of type } i}\right>
\end{equation}
\noindent and use the mass action laws to determine the association constants. From the association probabilities, we can compute the association constant using the mass action laws, which then determines all the parameters for our theory. Form our analysis of the 21~m water-in-LiTFSI, we found $\lambda = 0.2527$ and $v_0 = 21.744$ \AA$^3$. In addition, we used the following parameters $\epsilon_r = 10.1$, and $P$ = 4.995 Debye~\cite{mceldrew2018}.

\section{Results}

In this section, we validate our proposed EDL theory for 21~m LiTFSI WiSE~\cite{mceldrew2021ion,Markiewitz2025} at negative and positive electrodes against atomistic MD simulations~\cite{mceldrew2018}, which has not been achieved before. Notably, this is because 21~m is in the gel regime~\cite{mceldrew2021ion,Markiewitz2025}, and the system of equations becomes challenging to solve. Previously, in Ref.~\citenum{Markiewitz2025} the polynomial formulation was investigated. This worked well for the 12~m and 15~m LiTFSI, which is why we used this method. However, it was challenging to solve this system of equations for 21~m LiTFSI, which is why this comparison was not shown. 

From further testing different approaches for the 21~m LiTFSI, we found that overall the Boltzmann closure relations are slightly more numerically well behaved at higher concentrations. As such, we investigated alternative forms of the closure relationship, such as the new one outlined in the Theory Section. However, we found that the hydrated cation Boltzmann closure was more robust than other approaches, as as Eq.~\eqref{eq:close_cat_new}. Therefore, we stick to this formulation here. Note that these Boltzmann closure relations are formally derived in the pre-gel regime, and additional terms from the gel could be included, but it has previously been found these have a small effect~\cite{Goodwin2022EDL}. To further help with numerical solutions and comparisons to experiments, we introduced a consistent charge rescaling parameter~\cite{Goodwin2017a}, which makes the agreement more qualitative than in Ref.~\citenum{Markiewitz2025} for the 12~m and 15~m LiTFSI WiSEs. We used the largest value of $\alpha=0.3$~\cite{Goodwin2017a} that would allow us to solve the system of equations. Here we use this value for all calculations. 

\subsection{Negative EDL}

\begin{figure}[h!]
 \centering
 \includegraphics[width= 1\textwidth]{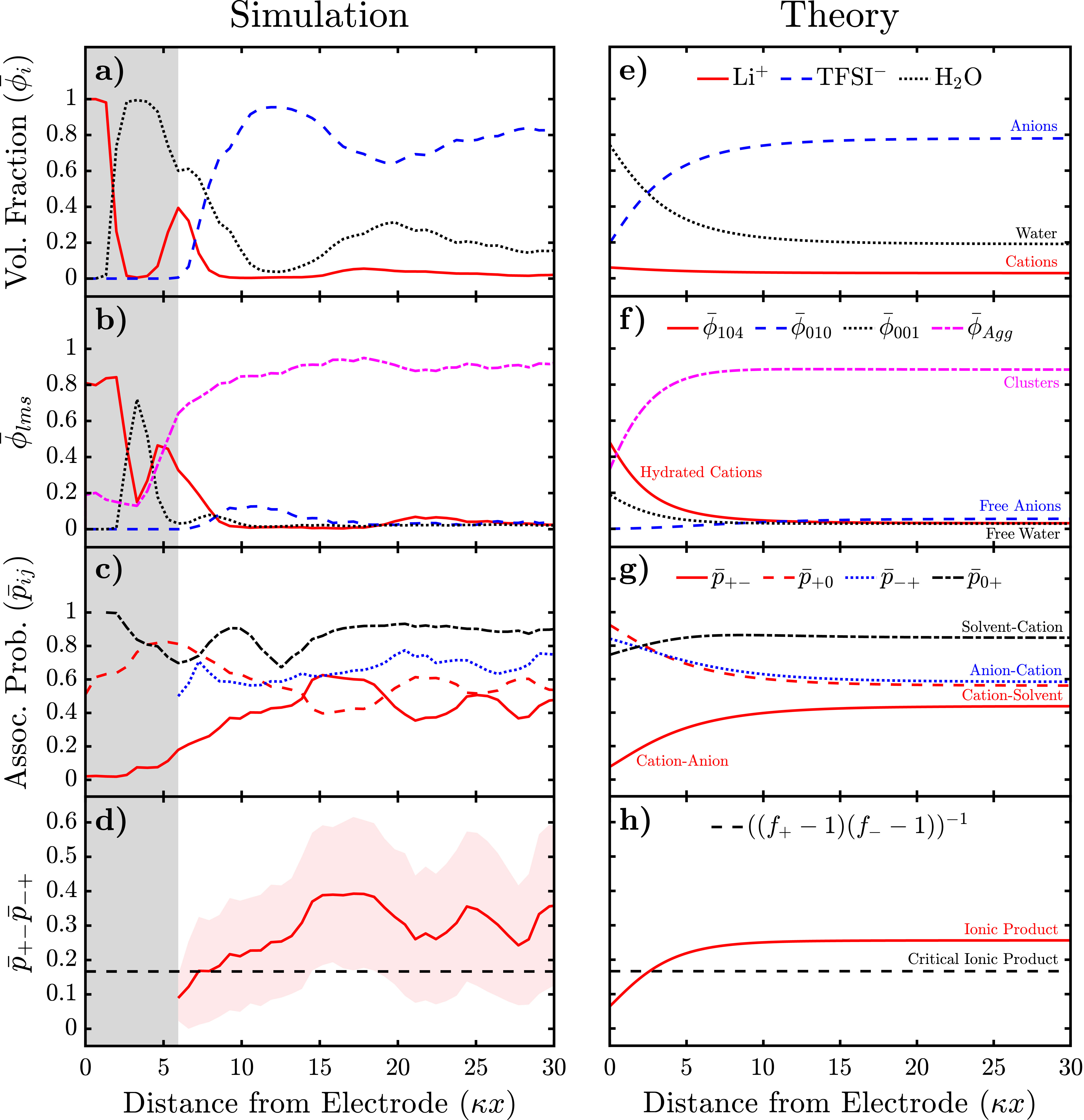}
 \caption{EDL of 21~m WiSEs at negative electrode ($\sigma$ = -0.2 C/m$^2$) as a function from the interface, in dimensionless units, where $\kappa$ is the inverse Debye length, from simulations (a-d) and theory (e-h). In the former, the grey region indicates the minimum distance from the electrode at which a species was never found. a,e) Volume fraction of each species ($\bar{\phi}_i$). b,f) Volume fractions of hydrated cations (Simulation $\bar{\phi}_{10x}$ \& Theory $\bar{\phi}_{104}$), free anions ($\bar{\phi}_{010}$), free water ($\bar{\phi}_{001}$), and aggregates and gel, if present ($\bar{\phi}_{Agg}$). c,g) Association probabilities ($\bar{p}_{ij}$). d,h) Product of the ionic association probabilities, $\bar{p}_{+-}\bar{p}_{-+}$, where the dashed line indicates the critical line for gelation and the red shaded region is the standard deviation.}
 \label{fig:neg}
\end{figure}

In Fig.~\ref{fig:neg}, we display our comparison between theory and simulation for the anode of 21m WiSE, where the grey region in the simulations denotes where not all species can be detected in the MD simulation. This demarcates the onset of where comparison to predictions in the theory should break down and where surface effects can become important (such as blocking of association sites and interactions with the electrode)~\cite{Markiewitz2025}. Hereon out, we refer to this as the Helmholtz region~\cite{GoodwinHelm2025}.

First we compare the volume fractions of each species in the EDL, $\bar{\phi}_i$, as seen in Figs.~\ref{fig:neg}(a)\&(e) for the simulation and theory, respectively. In the simulation, we find oscillations of $\bar{\phi}_-$ in the diffuse part of the EDL, with a strong decay towards the Helmholtz region, with the anions being completely depleted in this region. On the other hand, we find $\bar{\phi}_+$ and $\bar{\phi}_0$ have small oscillations in the diffuse EDL, but they accumulate strongly before the Helmholtz region. Moreover, we find two distinct layers of Li$^+$, one being right at the interface of the electrode in the Helmholtz region, and the other at the boundary of the Helmholtz region with the diffuse EDL, with a substantial water layer between them. The theory predictions are able to qualitatively capture these changes before the Helmholtz region, i.e., the diffuse EDL, but the theory cannot capture the oscillations and layering which occur, since it is a local density approximation. Specifically, the theory predicts a monotonic decay of anions and a monotonic increase of cations and water.

Next we compare the volume fractions of some chosen clusters, $\bar{\phi}_{lms}$, see Figs.~\ref{fig:neg}(b)\&(f), which allows us to understand the changes in species volume fractions in more detail. In the MD simulations, the volume fraction of free anions, $\bar{\phi}_{010}$, remains small throughout the EDL and bulk region, indicating most anions exist in aggregates. Similarly, there's a small volume fraction of hydrated cations and water in the bulk and diffuse EDL, but there's a notable increase in them towards the interface. We find that the two layers of Li$^+$ previously described are largely composed of hydrated cations, while the water layer between them is mostly free water. We also quantify the volume fraction of remaining aggregates and gel phase, calculated from $\bar{\phi}_{Agg} = 1 - \sum_x\bar{\phi}_{10x} - \bar{\phi}_{010} - \bar{\phi}_{001}$. In the bulk and diffuse EDL, the aggregates/gel are in majority, and only strongly dissipate when in the Helmholtz region, indicating that the aggregates/gel must be contributing substantially to the screening of the electrode. Again, the theory qualitatively captures these changes, in a smooth way, up to the Helmholtz region in the MD simulations. 

Finally we discuss how the association probabilities change within the EDL, which as far as we are aware no other theory is able to predict. In Figs.~\ref{fig:neg}(c)\&(g) we show $\bar{p}_{ij}$ from simulation and theory, respectively. In the diffuse region of the simulations there are oscillations in $\bar{p}_{+0}$ (the probability of cations associating to water), which transition into a steady increase towards the Helmholtz region. In contrast, $\bar{p}_{0+}$ remains approximately constant, with a slight decrease just before the Helmholtz layer, albeit with some fluctuations. In the theory, $\bar{p}_{+0}$ steadily increases in the EDL, but $\bar{p}_{0+}$ steadily decreases. Therefore, our theory has some success in describing these changes in coordination environments.

The cation-anion association probability, $\bar{p}_{+-}$, decreases substantially in the EDL from the simulations, while the anion-cation association probability, $\bar{p}_{-+}$, remains roughly constant in the simulations. Our theory also predicts $\bar{p}_{+-}$ to decrease in the EDL, but for $\bar{p}_{-+}$ to increase to a lesser extent. Our theoretical approach describes well the cation coordination environment, but struggles to predict the anion coordination environment. 

By taking the product of these probabilities, $\bar{p}_{+-}\bar{p}_{-+}$, we can determine bond percolation for the Bethe lattice that the aggregates exist on. If $\bar{p}_{+-}\bar{p}_{-+} > [(f_+ - 1)(f_- - 1)]^{-1}$, there is a percolating ionic network, i.e. a gel phase. In bulk 21~m LiTFSI WiSE, the percolating ionic network is known to exist, with the gel-point being determined to be close to 18m in LiTFSI (from the bond percolation criteria). In the simulations, as displayed in Fig.~\ref{fig:neg}(d), we see in the bulk and diffuse EDL that a gel phase exists, but close to the Helmholtz region the network is destroyed as this product dips below the critical value, which corresponds well to the reduction in aggregates/gel observed in Fig.~\ref{fig:neg}(b). The theory, shown in Fig.~\ref{fig:neg}(h), is able to capture the changes in the percolating ionic network well, where we predict a gel phase in the bulk and diffuse EDL, but close to the interface the criteria drops below the critical value.

Overall, in Fig.~\ref{fig:neg} we have demonstrated the overall changes in composition and coordination environments of 21~m LiTFSI WiSE EDL at negative electrodes, and found reasonable agreement with the the theory and simulations. As previously discussed, the largest differences between these methods occurs in the Helmholtz region, where not all species are present and surface effects become important. In Ref.~\citenum{GoodwinHelm2025}, it was suggested that an approach to describe the differences in the Helmholtz region was from the interface blocking association sites and interacting with them, manifesting through an apparent reduction in the functionalities, $f_i$. If we propose that $f_+$ changes from 4 to 3 in the Helmholtz region, we might expect to observe a decrease in $\bar{p}_{+j}$ in the simulations in this region, as a value of 4 was used in Fig.~\ref{fig:neg}, which is what we can see for $\bar{p}_{+j}$. 

\begin{figure}[h!]
     \centering
     \includegraphics[width= 1\textwidth]{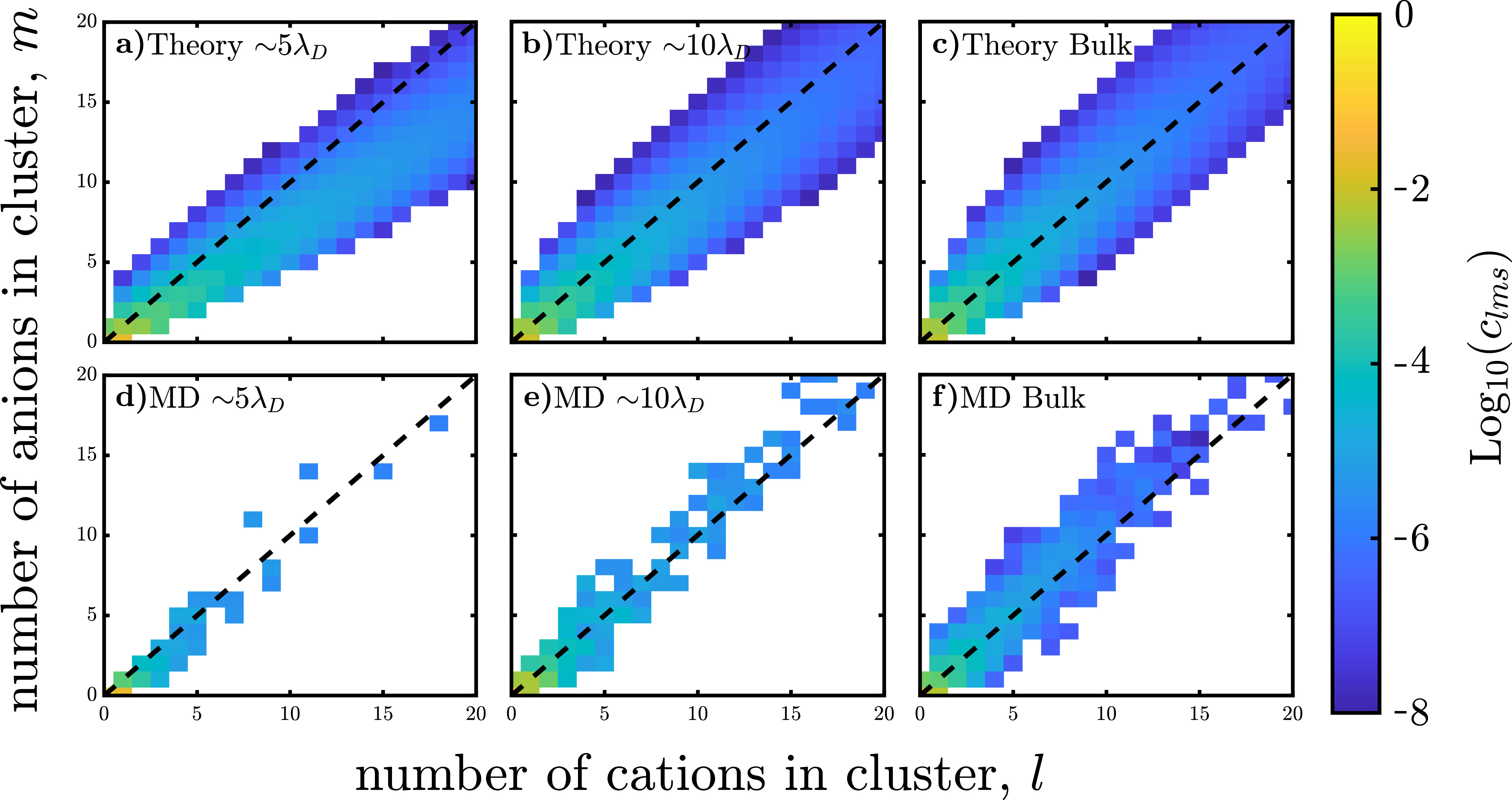}
     \caption{Cluster distribution of 21~m water-in-LiTFSI near the negative electrode ($\sigma$ = -0.2 C/m$^2$) at different distances from the interface, from theory (a-c) and simulations (d-f).}
     \label{fig:comp_n}
\end{figure}

Our approach also allows us to investigate the cluster distribution, $\bar{c}_{lms}$, in different regions of the EDL, which we display in Fig.~\ref{fig:comp_n} from simulation and theory. Note that we use the stick-cation approximation, so we only need to plot $\bar{c}_{lm}$. In the bulk, we observe an MD cluster distribution has a slightly negative bias to the aggregates, since $f_+ > f_-$, with clusters containing up to 12 cations and 12 anions, as seen in Fig.~\ref{fig:comp_n}(f). The theory has a similar distribution, see  Fig.~\ref{fig:comp_n}(c), but with a longer tail in the distribution. In the diffuse EDL, at $\sim10\lambda_D$ from the interface, we find that the MD cluster distribution has increased concentrations of some larger clusters, consistent with the observation that the gel phase is being dismantled closer to the interface, as displayed in Fig.~\ref{fig:comp_n}(e). For small clusters ($l < 5$), there is a positive bias of these aggregates, but for larger aggregates there's a negative bias. The theory captures the extended cluster distribution, but predicts a positive bias for all aggregates, shown in  Fig.~\ref{fig:comp_n}(b), indicating a breakdown in our assumed cluster distribution. Finally, just outside the Helmholtz region ($\sim5\lambda_D$), our simulations [Fig.~\ref{fig:comp_n}(d)] show a diminished cluster distribution with positive bias, demonstrating the screening ability of these aggregates. The theory, see Fig.~\ref{fig:comp_n}(a), has a strongly positive bias for the charge of clusters and the cluster distribution has grown relative to the diffuse EDL, which can be attributed to the theory over-predicting the stability of the gel phase $5\lambda_D$ from the interface.

\subsection{Positive EDL}

\begin{figure}[h!]
     \centering
     \includegraphics[width= 1\textwidth]{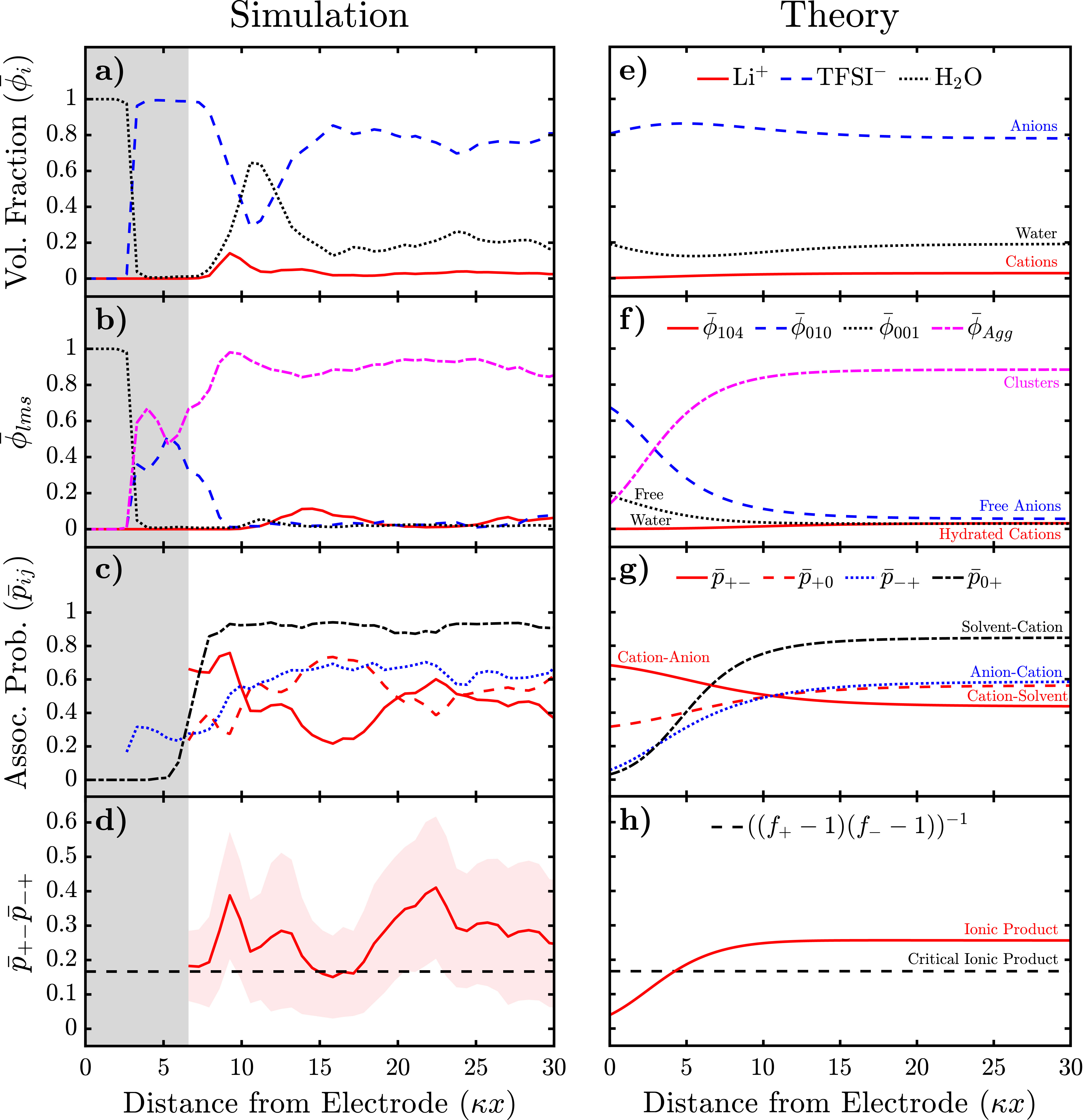}
     \caption{EDL of 21~m WiSEs at positive electrode ($\sigma$ = 0.2 C/m$^2$) as a function from the interface, in dimensionless units, where $\kappa$ is the inverse Debye length, from simulations (a-d) and theory (e-h). In the former, the grey region indicates the minimum distance from the electrode at which a species was never found. a,e) Volume fraction of each species ($\bar{\phi}_i$). b,f) Volume fractions of hydrated cations (Simulation $\bar{\phi}_{10x}$ \& Theory $\bar{\phi}_{104}$), free anions ($\bar{\phi}_{010}$), free water ($\bar{\phi}_{001}$), and aggregates and gel, if present ($\bar{\phi}_{Agg}$). c,g) Association probabilities ($\bar{p}_{ij}$). d,h) Product of the ionic association probabilities, $\bar{p}_{+-}\bar{p}_{-+}$, where the dashed line indicates the critical line for gelation, with the shaded region denoting its standard deviation.}
     \label{fig:pos}
\end{figure}

In Fig.~\ref{fig:pos}, we compare the EDL predictions from MD simulations and theory for the positive (cathodic) electrodes of 21m LiTFSI WiSE. We display the changes of volume fractions in the EDL in Fig.~\ref{fig:pos}a)\&e), from MD and theory, respectively. The simulations predict $\bar{\phi}_-$ to have a pronounced dip in the diffuse layer, before saturating just outside the Helmholtz region. Right at the interface, however, the anions are completely depleted and there is a substantial water layer. There is also a large accumulation of water in the dip in anion volume fraction in the diffuse EDL. Interestingly, there is a moderate increase in the Li$^+$ volume fraction between the dip in $\bar{\phi}_-$ and the peak in $\bar{\phi}_0$ (at $x \approx 9\lambda_D$), before it is completely suppressed in the Helmholtz region. The theory predicts more modest changes in the EDL. Cations are slowly, monotonically suppressed towards the interface. In contrast, the anions are initially enriched relative to the bulk, before being depleted closer to the interface, reflecting the MD simulations qualitatively. Similarly, the water is initially depleted in the diffuse layer, before enriching near the interface, also in the non-monotonic dependence observed in the simulations.

Further inspecting the common clusters and free species in the EDL reveals more insights into these changes in volume fraction as shown in Fig.~\ref{fig:pos}b)\&f). In the MD, we observe a large increase in free anions, $\bar{\phi}_{010}$, around the boundary of the Helmholtz region, which corresponds to the saturation of $\bar{\phi}_-$ previously described. Similarly, the water layer right at the interface is solely free water. The hydrated cations do not accumulate in the positive EDL. However, we do observe a non-monotonic dependence of the aggregate/gel volume fraction as the electrode interface is approached. In the diffuse EDL, there is a large peak which corresponds well to the increase in Li$^+$, indicating the cations are mainly bound up in aggregates. Moving towards the Helmholtz region, the aggregates/gel is suppressed, but they remain substantial in the middle of the Helmholtz region. The theory agrees reasonably well with these simulations, although the finer, non-monotonic details are missed. We find a strong suppression in hydrated cations and clusters/gel, and an accumulation of free anions and free water. 

Next we inspect the changes in association probabilities, as seen in Fig.~\ref{fig:pos}c)\&g). From the simulations, we find $\bar{p}_{+0}$ fluctuates in the bulk and diffuse region before being suppressed near the Helmholtz region. Similarly, we find $\bar{p}_{0+}$ is strongly suppressed to 0 in the Helmholtz region indicating that water prefers to be in the free state. These changes in association probabilities are captured qualitatively by the theory. We find $\bar{p}_{+-}$ also fluctuates in the bulk/diffuse EDL, before increasing slightly towards the Helmholtz layer. While the anion-cation association probability decreases in the EDL. The theory is able to qualitatively reproduce these changes in the ionic association probabilities.

In Fig.~\ref{fig:pos}d)\&h), we investigate where the percolating ionic network survives in the EDL. From the simulations, we find the bulk the gel phase exists, but moving towards the diffuse and Helmholtz regions we find large fluctuations, where it is almost suppressed several times. Within the Helmholtz region no gel phase exists, since we know this is dominated by free water and free cations, with some clusters. We find the theory predicts a monotonically decreasing percolation probability, which just dips below the critical condition at the interface, indicating the gel phase is destroyed there.  

\begin{figure}[h!]
     \centering
     \includegraphics[width= 1\textwidth]{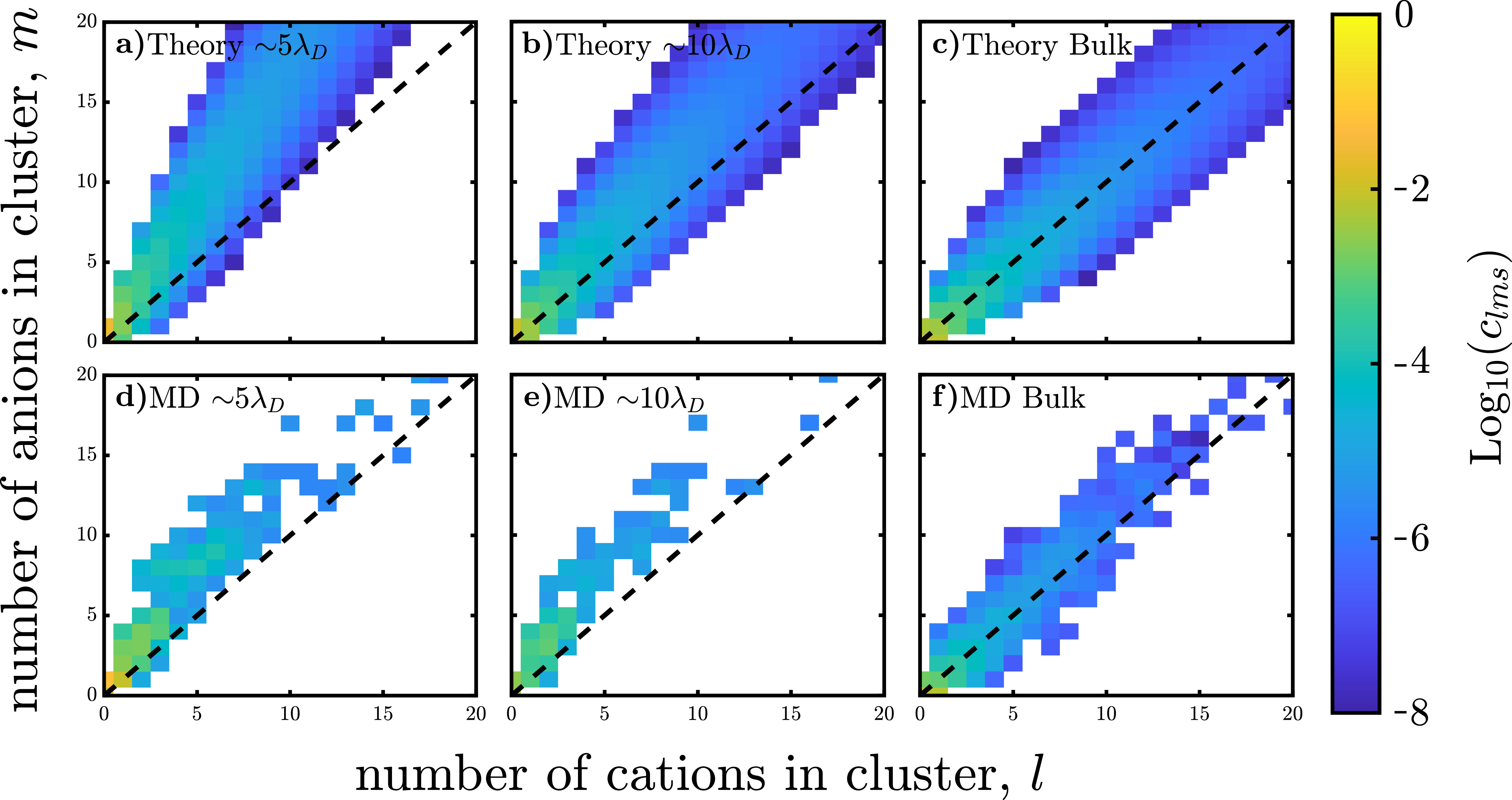}
     \caption{Cluster distribution of 21m water-in-LiTFSI near the positive electrode ($\sigma$ = 0.2 C/m$^2$) at different distances from the interface, from theory (a-c) and simulations (d-f).}
     \label{fig:comp_p}
\end{figure}

In Fig.~\ref{fig:comp_p} we further inspect the cluster distribution, from theory and simulation, at various positions in the EDL at positive electrodes. We will not recap the bulk as it was previously described. In the diffuse layer $\sim10\lambda_D$ from the interface, the simulations predict a negatively biased cluster distribution. The simulations predict that the concentrations of the larger clusters increase relative to the the bulk, but some of the larger aggregates are suppressed relative to the bulk. The theory predicts that the cluster distribution has a negative bias and larger clusters exist than the bulk. At $\sim5\lambda_D$ similar observations are found from both theory and simulation.

\subsection{Thermodynamic Stability and Interphase Formation}

Having further validated that our theory captures the trends in composition and coordination environments in the EDL of 21~m LiTFSI WiSE~\cite{Markiewitz2025}, we turn to using this theory to understand thermodynamic effects that influence stability and interphase formation of these electrolytes~\cite{mceldrew2021ion,mceldrew2018,phelan2026linking}. We achieve this through predicting the changes in thermodynamic activity in the bulk and EDL, and changes in concentration within the EDL~\cite{mceldrew2021ion,mceldrew2018,Markiewitz2025}.

We start by deriving expressions for the thermodynamic activity in the bulk with the sticky-cation approximation~\cite{mceldrew2021ion}, based on our functional free energy formalism~\cite{Markiewitz2025}. Generally, the activity of a species, $a_i$, can be expressed as 

\begin{equation}
   \ln a_i = \beta (\mu_i - \mu_i^{\theta}),
   \label{eq:act}
\end{equation}

\noindent where $\mu_i$ is the chemical potential and $\mu_i^{\theta}$ is the reference state chemical potential. Using this definition, we have for the bulk, sticky-cation approximation (neglecting gel terms, as we did for the closure relations) activity of each species

\begin{equation}
\ln\left(a_{+}\right)=\ln\left(\frac{\phi_+}{\phi_+^\theta}\right) + f_+\ln\left(\frac{p_{+0}}{\psi_0[1 - p_{0+}]}\frac{\psi_0^\theta[1 - p_{0+}^\theta]}{p_{+0}^\theta} \right),
\label{eq:Cat_act_b}
\end{equation}

\begin{equation}
    \ln\left(a_{-}\right)= \ln\left(\frac{\phi_-}{\phi_-^\theta}\right) + f_-\ln\left(\dfrac{1-p_{-+}}{1-p_{-+}^\theta}\right),
\label{eq:An_act_b}
\end{equation}  

\begin{equation}
\ln\left(a_{0}\right)= \ln\left(\dfrac{\phi_0[1-p_{0+}]}{\phi_0^\theta[1-p_{0+}^\theta]}\right).
\label{eq:W_act_b}
\end{equation}

\noindent Note that with how we establish the chemical equilibrium, the chemical potential of cations (same logic applies to anions and water) in an aggregate is equal to the free cation, and the free cation chemical potential does not depend on the cluster rank, so it is the convenient choice ($\mu_{lms}^+ = \partial \mu_{lms}/\partial l = \mu_{100} = \mu_+$)~\cite{mceldrew2020theory,mceldrew2021ion}. In addition, we can also calculate the activity of certain species, such as the hydrated cation

\begin{equation}
\label{eq:Cat_Hyd_act_B}
\ln\left(a_{10f_+}\right) = \ln\left(\frac{\phi_+}{\phi_+^\theta}\right) + f_+\ln\left(\frac{p_{+0}}{p_{+0}^\theta}\right), %- (\xi_++f_+)\Lambda 
\end{equation}

\noindent which is an important species in WiSEs. Note that in our theory, the activity, which represent the effective concentrations of species, is given by the concentration of free species~\cite{mceldrew2020theory}. This provides a simple and direct link, where all of the non-ideal changes are occurring through the associations.

In Eqs.~\eqref{eq:Cat_act_b}-\eqref{eq:W_act_b}, we take the reference state to be infinite dilution for the electrolyte. This corresponds to $\phi_0^\theta = 1$, $p_{+-}^\theta = p_{-+}^\theta = p_{0+}^\theta = 0$, and $p_{+0}^\theta = 1$. We can use this reference state explicitly, but divergences remain. In practice the standard is take a small non-zero concentration of the electrolyte as the reference, which removes these divergences~\cite{mceldrew2021ion}. 

\begin{figure}
    \centering
    \includegraphics[width=0.5\linewidth]{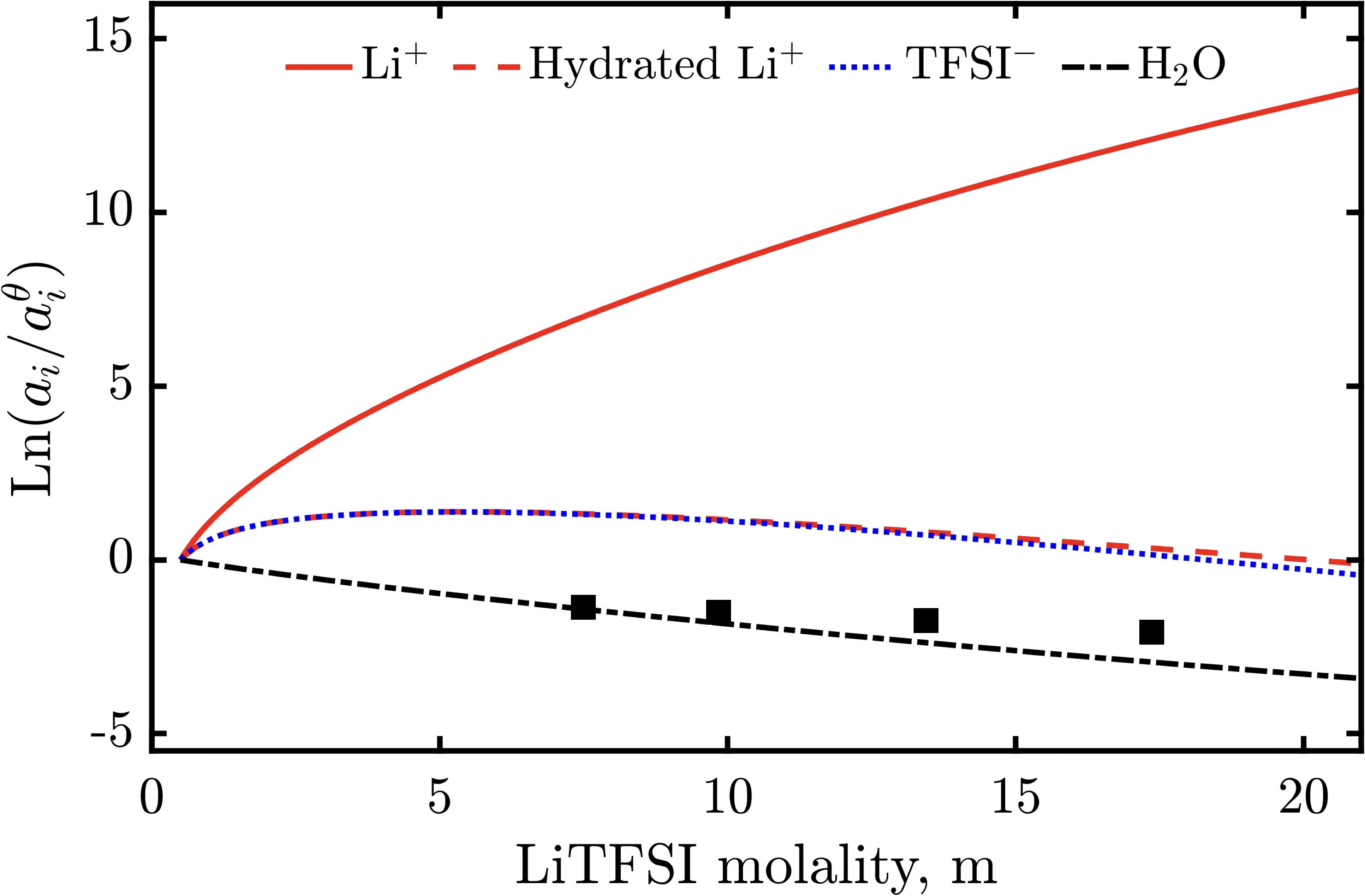}
    \caption{Activity of ions, water and hydrated cations in the bulk as a function of molality, computed with the sticky cation approximation for LiTFSI. The reference concentration was taken to be 0.5~m. Experimental data, reproduced and adapted from Ref.~\citenum{Han2021WiSE}, is referenced to align with the calculated activity at the lowest concentration.}
    \label{fig:act_bulk}
\end{figure}

In Fig.~\ref{fig:act_bulk}, we show the changes in activity as a function of salt concentration from these equations, for water, cations, anions and hydrated cations referenced against the 0.5~m state. Consistent with previous findings~\cite{mceldrew2021ion}, we see the water activity decreases with salt concentration, the anion activity initially increases with salt concentration before decreasing again at higher concentrations, and the Li$^+$ activity monotonically increases with salt concentration~\cite{mceldrew2021ion,Han2021WiSE}. In addition, in Fig.~\ref{fig:act_bulk} we show the activity of water derived from the experiments of Han \textit{et al.}~\cite{Han2021WiSE}, where we have aligned the experiments with the theory predictions at the lowest concentration. In the experiments, the water activity also decreases with increasing salt concentration, but to a lesser extent than we predict with out theory. We find the activity of hydrated cations tracks the activity of the anions. 

\begin{figure}
    \centering
    \includegraphics[width=0.5\linewidth]{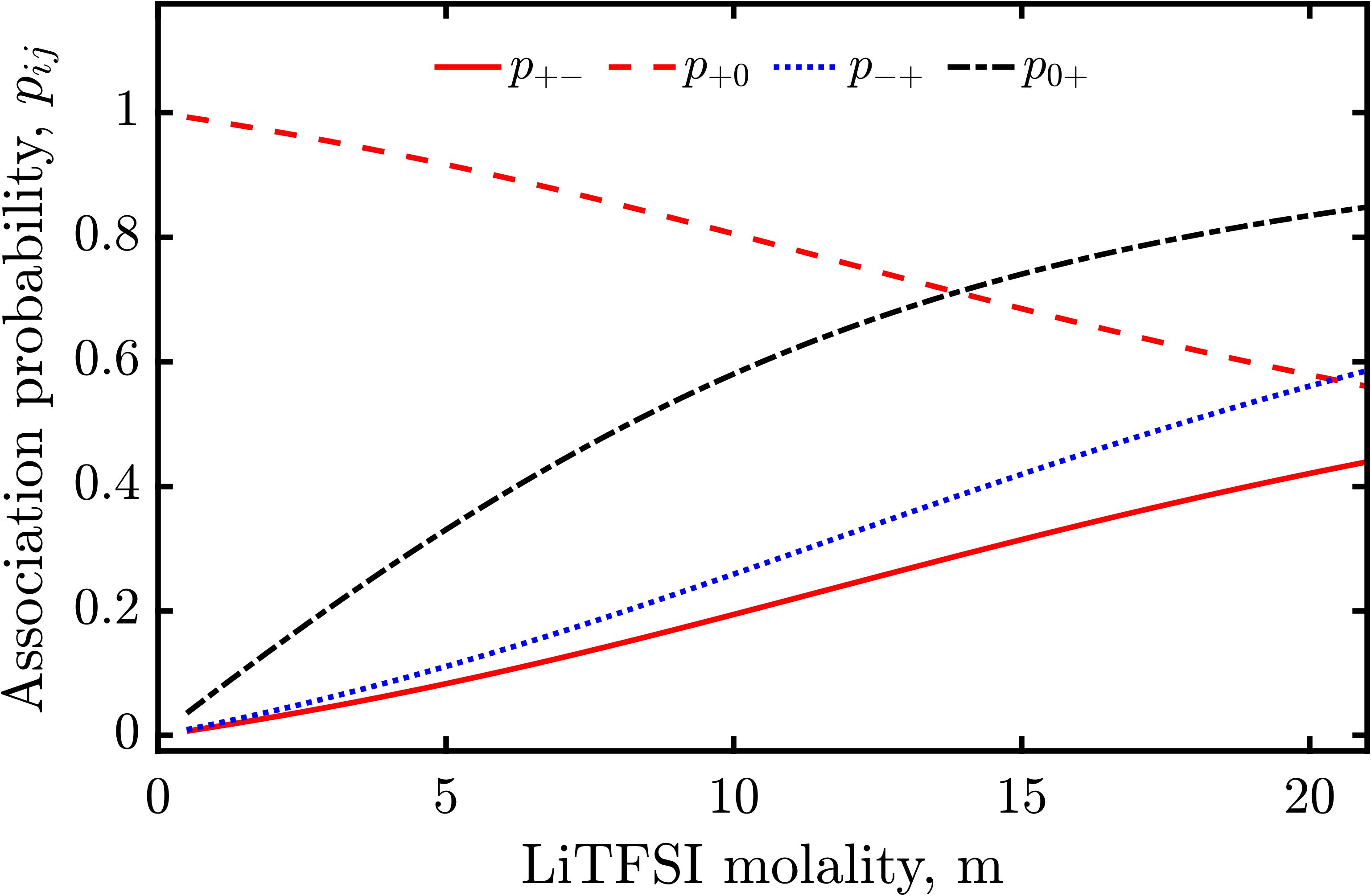}
    \caption{Association probabilities as a function of salt molality, from 21m down to 0.5m, computed with the sticky-cation approximation for LiTFSI.}
    \label{fig:bulk_p}
\end{figure}

To understand these changes in activity, all we need to know is how the association probabilities change with salt concentration, which provides the direct link of how activity depends on coordination environments as well as how the volume fractions of species change with salt concentration. In Fig.~\ref{fig:bulk_p}, we show how the association probabilities change with salt concentration. Using this information, we can determine that the anion activity is non-monotonic because of the increase in $p_{-+}$ dominates over the increase in $\phi_-$ at large salt concentrations. The cation activity monotonically increases because $p_{+0}/\psi_0(1 - p_{0+})$ also increases with salt concentration. The activity of water decreases because of a decreasing $\phi_0$, but also because of an increasing $p_{0+}$. 

To understand why the hydrated cation activity follows that of the anion, we expand it in the smallness of $p_{-+}$ and $p_{+-}$ at low concentrations to give $\ln\left(a_{10f_+}\right) \approx f_+ p_{+-}$, and $\ln\left(a_{010}\right) \approx f_-p_{-+}$, which we can see must hold from the conservation of associations. According to hydration theory~\cite{Daghetti1993}, the difference between the activity of cations and anions can solely be attributed to their different solvation environments. Our observation that the hydrated cation activity is close to the anion activity suggests the hydration theory~\cite{Daghetti1993} could be contained within, or be a limit of, our more general theory.

These changes in activity are a key part of the changes in thermodynamic stability of the electrolyte, and its ability to form a stable interphase. The changes in activity can be directly linked, through the Nernst equation, to changes in redox potentials of reactions which can occur to form the interphases, or parasitic reactions. 

Let us consider the SEI reaction from Ref.~\citenum{steinruck2020interfacial}, where the TFSI$^{-}$ undergoes reduction and removal of flouride, which combines with the Li$^+$, to form LiF
\begin{equation}\label{eq:lif_formation}
    \text{Li}^++(\text{R-F}_3)^- + \text{e}^- (\text{sec. electrons}) \rightleftharpoons \text{LiF}+ (\text{R-F}_2)^- 
\end{equation}

\noindent where R=CF$_3$-SO$_2$-N-SO$_2$-C. The change in the equilibrium reduction potential is given by the Nernst equation of this reaction
\begin{align}
    E - E^{\theta} = \dfrac{k_BT}{e}\ln\left(\frac{a_{\text{Li}^{+}}a_{(\text{R-F}_3)^-}}{a_{\text{LiF}}a_{(\text{R-F}_2)^-}}\right),
\end{align}

\noindent Taking the activity of the products are ideal, i.e.,  $a_{\text{LiF}} = 1$ and $a_{(\text{R-F}_2)^-} = 1$, we have for the change in equilibrium redox potential

\begin{align}
    E-E^{\theta}=\dfrac{k_BT}{e} \ln \left(a_{\text{Li}^{+}}a_{(\text{R-F}_3)^-}\right) = k_BT \ln (a_{+}a_{-})
\end{align}

\noindent Therefore, increasing the salt concentration from 0.5~m to 2~ m leads to a positive shift of approximately 0.35~eV in electrochemical reactions involving Li$^+$, such as LiF formation in Eq.~\eqref{eq:lif_formation}. Similarly, other Li-containing salts formed in the SEI/CEI (Li$_x$PO$_y$F$_z$, Li$_2$O, Li$_2$CO$_3$, and LiOH)~\cite{phelan2024role, bjorklund2022cycle, rinkel2020electrolyte}, as well as their associated decomposition reactions\cite{rinkel2022two}, will undergo positive shifts in their electrochemical reduction potentials as the concentration increases from 0.5~m to 21~m due to changes in Li activity. Moreover, our theory shows that variations in Li$^+$ activity play a significantly larger role in shifting these reduction potentials than changes in TFSI$^-$ activity, the latter contributing a comparatively small negative shift of approximately -0.1 V over the same concentration range. These shifts lead to significant differences in the chemical composition of the interphases formed at 0.5~m and 21~m.

Next we consider the salt stabilization of water. At positive interfaces, suppression of the oxygen evolution reaction has been reported. Considering the oxidation of water, we have
\begin{equation*}
    2\text{H}_2\text{O} \rightleftharpoons \text{O}_2 + 4\text{H}^+ + 4e^-
\end{equation*}

\noindent The Nernst equation for oxygen evolution reaction is

\begin{equation}
        E-E^{\theta}= -\dfrac{k_BT}{4e} \ln \left(\dfrac{a_{H_2O}^2}{a_{H^+}^4a_{O_2}}\right).
\end{equation}

\noindent Assuming the activity of products are ideal, $a_{H^+} \approx 1$ (i.e., neglecting the role of pH) and $a_{O_2} \approx 1$, we have

\begin{equation}
        E-E^{\theta}= -\dfrac{k_BT}{2e} \ln \left(a_{H_2O}\right) = -\dfrac{k_BT}{2e}\ln (a_{0}).
\end{equation}

\noindent This states changing the salt concentration form 0.5~m to 21~m shifts the oxidation potential of water by 5~mV. Therefore, it becomes slightly less thermodynamically favourable to oxidise water with increasing salt concentration. This indicates improvements in electrochemical stability for the water mainly arise from kinetic contributions as opposed to thermodynamic contributions at more positive potentials. 

In a similar way, consider the reduction of water at negative potentials

\begin{equation*}
    2\text{H}_2\text{O} + 2\text{e}^- \rightleftharpoons \text{H}_2 + 2\text{OH}^-
\end{equation*}

\noindent The Nernst equation for this reaction is
\begin{equation}
        E-E^{\theta}=\dfrac{k_BT}{2e} \ln \left(\dfrac{a_{H_2O}^2}{a_{OH^-}^2a_{H_2}}\right).
\end{equation}

\noindent Again, assuming the activity of products are ideal, $a_{OH^-} \approx 1$ (i.e., neglecting the role of pH) and $a_{O_2} \approx 1$, we arrive at

\begin{equation}
        E-E^{\theta}=\dfrac{k_BT}{e} \ln \left(a_{H_2O}\right) = \dfrac{k_BT}{e}\ln (a_{0}).
\end{equation}

\noindent Increasing the salt concentration from 0.5 m to 21 m shifts the reduction potential of water by -10 mV, making water reduction slightly less thermodynamically favorable at higher salt concentrations. This shift contributes, in part, to the apparent expansion of the electrochemical stability window. These results demonstrate that increasing salt concentration appreciably alters SEI composition by promoting the formation of Li salts, while thermodynamically suppressing the reduction of water. While the thermodynamic stability at positive potentials is slightly increased, kinetic limitations of the OER likely dominate resulting in the improved oxidative stability observed in WiSEs. %\textcolor{red}{citing jaspers paper here would also be nice}

%https://pubs.acs.org/doi/pdf/10.1021/acscatal.3c04255?ref=article_openPDF
Next, we look into the thermodynamic effects which can alter reaction kinetics at the interface~\cite{Bazant2009a,Fedorov2014}. In expressions for the current densities, there is often a linear dependence on the concentration of the species reacting at the interface~\cite{Fedorov2014}. The seminal example of which is Butler-Volmer reaction kinetics~\cite{Fedorov2014}. Moreover, this intuitively makes sense, since what can react (and form the interphase or evolve gasses) is what is at the interface between the electrode/interphase and the electrolyte, where the electron transfer processes can occur. Therefore, provided we are in a situation where transport limitations are not strong, such that an equilibrium theory can be used, looking at the concentration of key species in the EDL, especially right at the interface, can provide insight into changes in these redox reactions. For the reactions previously discussed, we consider how the concentration of reactants at the interface could influence the rates of these reactions. 

For the formation of the inorganic SEI, we can inspect the surface concentrations of Li$^+$ and TFSI$^-$~\cite{steinruck2020interfacial}. As previously discussed, we found a large concentration of Li$^+$ right at the interface of negative electrodes, followed by a layer of water, and at further distances the are significant TFSI$^-$ anions. Further inspection showed that much of the Li$^+$ and water were there in the form of hydrated Li$^+$, but there was also a significant fraction of clusters. Since these clusters will involve Li$^+$-TFSI$^-$ at the interface, there should not be any kinetic reasons to limit the formation of LiF, and in fact, these cluster should facilitate this reaction. At lower concentrations, as previously studied in Ref.~\citenum{Markiewitz2025}, the interface was significantly more populated with hydrated Li$^+$ and there was an absence of clusters at the interface. Therefore, with increasing salt concentration, there is an increase in the thermodynamic stability of the electrolyte, but also the initial kinetics of the SEI formation should be more facile, which can also contribute to a longer-term kinetic passivation through the formation of this SEI~\cite{steinruck2020interfacial}. On the other hand, for the oxidation of water at $+$0.2 C/m$^2$ interfaces, we find substantial water right at the interface~\cite{steinruck2020interfacial}. While this is a large surface charge, as studied in Refs.~\citenum{mceldrew2018,Markiewitz2025}, at lower surface charges there can be a depletion of water, which would suppress the kinetics of the oxygen evolution reaction. 

Not only can we inspect the changes in concentration within the EDL, but we can also derive expressions for the activity within the EDL~\cite{Finney2021,Gebbie2023}. Again within the sticky-cation approximation, neglecting the gel terms and the electrostatic contributions to the electrochemical potential, we arrive at

\begin{equation}
\ln\left(\bar{a}_{-}\right) = \ln\left(\frac{\bar{\phi}_-}{\phi_-^\theta}\right) + f_-\ln\left(\dfrac{1-\bar{p}_{-+}}{1-p_{-+}^\theta}\right) - \xi_-\Lambda ,
\label{eq:An_act_EDL}
\end{equation}  

\begin{equation}
\ln\left(\bar{a}_{0}\right) = \ln\left(\dfrac{\bar{\phi}_0[1-\bar{p}_{0+}]}{\phi^\theta_0[1-p^\theta_{0+}]}\right) - \Lambda. 
\label{eq:Sol_act_EDL}
\end{equation}

\begin{equation}
\ln\left(\bar{a}_{+}\right)=\ln\left(\frac{\bar{\phi}_+}{\phi_+^\theta}\right) + f_+\ln\left( \dfrac{\psi_0^\theta\bar{p}_{+0} (1 - p_{0+}^\theta)}{\bar{\psi}_0  p_{+0}^\theta(1 - \bar{p}_{0+})}\frac{\sinh(\beta d|\nabla \Phi|)}{\beta d|\nabla \Phi|} \right) - \xi_+\Lambda
\label{eq:Cat_act_EDL}
\end{equation}

\noindent or equivalently,

\begin{equation}
\ln\left(\bar{a}_{+}\right)=\ln\left(\frac{\bar{\phi}_+}{\phi_+^\theta}\right) + f_+\ln\left(\dfrac{\psi_-^\theta\bar{p}_{+-} (1 - p_{-+}^\theta)}{\bar{\psi}_-  p_{+-}^\theta(1 - \bar{p}_{-+})}  \right) - \xi_+\Lambda,
\label{eq:Cat_act_EDL_II}
\end{equation}

\noindent similar to the activity expressions previously stated.

%https://wrap.warwick.ac.uk/id/eprint/190133/1/d1sc02289j.pdf
%\noindent Now these expressions have been referenced relative to the bulk and infinite dilution, which is why the electrostatic potential and Langevin terms remain, as well as the Lagrange multiplier. However, it is more typical to neglect these terms, since these largely cancel with the changes in activity in the EDL, owing to the equilibrium between the bulk and EDL. Therefore, hereonout, we use a version of the equations were these extra terms are neglected, and the equations are essentially the same as in the bulk, but where all quantities have been replaced by their EDL counterparts. 

\begin{figure}
    \centering
    \includegraphics[width=1\linewidth]{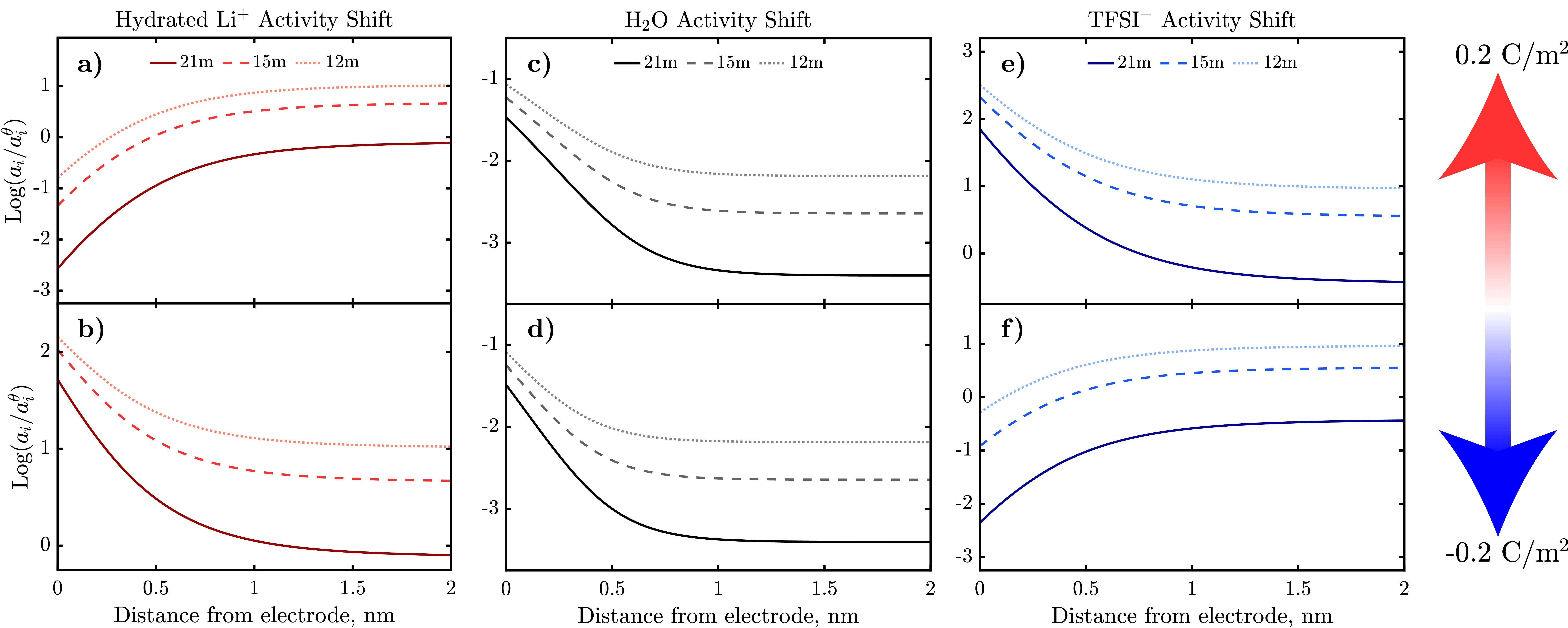}
    \caption{Activity of species within the EDL for as a function of distance from the electrode, at various concentrations, at the indicated surface charges with a temperature of 300~K.}
    \label{fig:EDL_act}
\end{figure}

In Fig.~\ref{fig:EDL_act}, we display how the activity changes within the EDL for 3 concentrations, for surface charges of $\pm0.2$Cm$^{-2}$. As expected from the previously described EDL profiles in Figs.~\ref{fig:comp_n}-\ref{fig:comp_p}, the activity of cations increases at negative electrodes, but decreases towards positive electrodes. This change is more pronounced with increasing concentration. Analogously, the anion activity decreases in the EDL of negative electrodes, but increases in positive EDLs, with larger effects for more concentration solutions. Whereas, we find the activity of water increases for both polarizations of the electrode, since water is attracted by electric fields~\cite{Markiewitz2025}. 

While we can determine the activity of species in the EDL~\cite{Finney2021}, which is commonly used terminology in the electrocatalysis community~\cite{Gebbie2023}, we note that this can lead to some confusion. In the Nernst equation, the electrochemical equilibrium between the oxidised and reduced species is considered in the bulk~\cite{mceldrew2021ion,phelan2026linking}. We also investigate the changes in composition in the EDL from establishing the equilibrium with the bulk, which is conceptually similar to the Nernst equilibrium. However, the Nernst equation considers the changes in chemical potential of the species with composition, while the EDL has a changing composition from the applied field, but the chemical potential of the species are equal to their bulk values~\cite{Goodwin2022EDL,Markiewitz2025}. Therefore, one must not think that the activity of species within the EDL can drastically change the redox potentials. The relevant quantities within the EDL are the concentrations of species, but these quantities sit in kinetic equations for these electron transfer processes, not for thermodynamic changes.

\section{Discussion}

Here, we presented a theory validated from MD simulations that predicts the structure of the EDL in 21~m LiTFSI WiSE~\cite{Markiewitz2025}, in addition to the thermodynamic activity of the species~\cite{mceldrew2021ion}, which allows for statements about changes in stability and reactivity towards interphase formation of the electrolyte. Our theory is simple, analytical and has at most one free fitting parameter ($\alpha$), with the handful of parameters it does have being easily obtained from MD simulations~\cite{mceldrew2020theory}. Moreover, it directly links changes in coordination environments to these changes in activity, therefore, confirming correlations between data that have previously been drawn~\cite{Suo2015}. We believe our approach has utility over others, such as the liquid-Madelung potential~\cite{takenaka2024liquid} or OPAS~\cite{mceldrew2021ion,phelan2026linking}, since only one MD simulation is required to parameterize the model, which can then make predictions over different compositions in the bulk and the EDL. 

Not only that, but our approach can be extended to other electrolytes readily. For example, we have also studied ionic liquids~\cite{mceldrew2020corr}, salt-in-ionic liquids~\cite{McEldrewsalt2021,markiewitz2024} and several conventional carbonate battery electrolytes~\cite{Goodwin2023,phelan2026linking}. In terms of battery electrolytes, Phelan \textit{et al.}~\cite{phelan2026linking} found similar trends in activity of salt and solvents. These predictions were validated from OPAS MD simulations, and also shifts in the Li$^+$/Li redox potential experimentally~\cite{phelan2026linking}. There, the liquid-Madelung potential was also compared~\cite{takenaka2024liquid}, where much worse agreement with MD OPAS and experimental measurements was found~\cite{phelan2026linking}.

Therefore, our formalism can be applied to a wide variety of electrolytes of interest in energy storage technologies. As described in the Theory Section, the presented version of the theory can only strictly be applied to an electrolyte where 2 components have $f_i > 2$, otherwise it is not known how the aggregates are connected~\cite{McEldrewPhD}. However, despite this, the mass action laws, conservation of associations, and expressions for the free concentrations of ions are independent of this approximation, and can be applied more generally to more complicated electrolytes with more components~\cite{McEldrewPhD,phelan2026linking}. These equations are

\begin{equation}
    \Gamma_{ij} \lambda_{ij} = \dfrac{p_{ij}p_{ji}}{(1 - \sum_{j'}p_{ij'})(1 - \sum_{i'}p_{ji'})},
\end{equation}

\begin{equation}
    \Gamma_{ij} = \psi_ip_{ij} = \psi_jp_{ji},
\end{equation}

\begin{equation}
    \phi_{f,i} = \phi_i\left(1 - \sum_{j'}p_{ij'}\right)^{f_i},
\end{equation}

\noindent where $\phi_{f,i}$ has been used to denote the free volume fraction of species $i$. While these equations allow for less restrictive functionalities to be used (any choice of functionality can be used), the electrolyte should still adhere to other assumptions of the theory (such as branched aggregates forming, instead of crystalline phases, and an association being well described by one association constant/environment~\cite{Tkachenko2025}), and needs to be well described by these approximations.

Moreover, assuming the species only interact through associations, we can use them in the activity expressions

\begin{equation}
    \ln(a_i) = \ln(\phi_{f,i}) = \ln\phi_i + f_i\ln\left(1 - \sum_{j'}p_{ij'}\right).
    \label{eq:act_gen}
\end{equation}

% but chemistries can be tuned in a way to reduce these association probabilities
\noindent With this, we can discuss more generally design principles of electrolytes to shift the activities in desired directions. To increase the activity of a species $i$, we can increase the concentration of $i$ or reduce $\sum_{j'}p_{ij'}$. While we have direct control over the composition of the electrolyte, how the species interact and coordinate is determined more by their chemistry. For example, in the context of high entropy electrolytes for Li-metal anodes, it was suggested that more components of a similar chemistry improves transport properties, but keeps the stability of the electrolyte~\cite{Cheol2023HE}. It is clear from Eq.~\eqref{eq:act_gen} and the discussion in Ref.~\citenum{Goodwin2023}, that the salt activity of a high entropy electrolyte wouldn't be altered, which is consistent with experimentally observations~\cite{Cheol2023HE}. With the generalised approach, we can also understand the many-anion high-entropy electrolytes~\cite{Wang2023HEE2}. Provided all anions interact with the cation with the same association constant, have the same functionality, and the overall volume fraction of anions is the same, there should then again be no net effect on activity. Any changes in these variables could, however, alter the predictions of activity, but it is likely their changes would be small (compared to the large changes over a wide range of salt concentrations). In those many anion high entropy electrolytes, similar SEI forming abilities were reported, which we can corroborate with our approach. In addition, improved transport properties and operating temperatures were reported~\cite{Wang2023HEE2}, but further investigation is required to comment on the design principles for transport properties. 

Moreover, the EDL theory developed is built on the equilibrium of free species~\cite{Goodwin2022EDL}. Therefore, a general closure relation can be stated

\begin{equation}
    \bar{\phi}_{f,i} = \phi_{f,i}F(\Phi,\nabla \Phi,\alpha),
\end{equation}

\noindent where $F(\Phi,\nabla \Phi,\alpha)$ is the response of the species to electrostatic potentials and fields. Note that an equivalent set of mass action laws and conservation of associations applies in the EDL~\cite{Goodwin2022EDL}. Therefore, the EDL can be investigated for more complicated electrolytes. 

What this approach lacks, however, is the microscopic insight into changes in association constants with composition~\cite{McEldrewsalt2021}. This could, however, be overcome empirically, and this simpler set of equations might provide more utility than the full complexity of the theory presented here.

In the context of interphase formation, we have several key thermodynamic insights, that can be widely applied to other electrolyte systems. However, the role of the electrode in our theory, and also simulation, is somewhat lacking, other than it holding the charge in the EDL. This was identified in Ref.~\citenum{GoodwinHelm2025}, where specifically the Helmholtz layer solvation environments of conventional battery electrolytes was studied. However, that approach was still mainly focused on the electrolyte properties, with the role of the electrode entering through a reduction in the apparent functionality of species (and a generic surface interaction potential). Therefore, to get more insight into the role of different electrodes, such as Li-metal vs graphite, these surface interactions need to be included more explicitly.

While we have discussed thermodynamic effects from our simple model which provides insight into interphase formation, it still cannot directly predict what is going to form in the SEI or CEI. The changes in activity correlate with changes in frontier orbitals~\cite{phelan2026linking}, but this does not provide an explicit link with interphase formation. Therefore, while our approach can provide guiding principles, to get chemically accurate predictions of interphase formation, atomistic approaches are necessary. Such approaches could be reaction networks~\cite{Wen2023net}, MLIPs, adiabatic energy surfaces from DFT~\cite{Abraham2025} for CIET~\cite{Fraggedakis2021,bazant2023unified}. While computationally challenging, these approaches can provide the chemical insight lacking in our approach. 

Finally, beyond insights into electrolyte stability and interphase formation, the EDL predictions here are also of fundamental importance, as many open questions remain in the context of the EDL of highly concentrated electrolytes~\cite{Espinosa2023rev,Goodwin2021Rev}. Anomalous underscreening, the observation of extremely long force decay lengths in surface force apparatus/balance (SFA/SFB) measurements, has puzzled the community for over a decade~\cite{Gebbie2013,Gebbie2015,Han2020,smith2016electrostatic,perez2017underscreening,Han2021WiSE,Zhang2024}. Extensive ion pairing, network formation~\cite{Gebbie2013} and ``holes''~\cite{perez2017underscreening} acting as the charge carriers have been proposed to explain these measurements. Our theory accurately includes ionic aggregates and gel in a consistent EDL framework, but we still only predict screening lengths marginally over the Debye length. Therefore, our theory does not support the assumptions of these measurements, i.e., that it is an equilibrium electrostatic effect. Recently, experimental evidence that these measurements are not at equilibrium have been reported~\cite{Cross2026}. Further investigating these effects is important, with a particular focus on the non-equilibrium changes in associations in the EDL~\cite{Zhang2024}.

Moreover, even still, the close range structure of the EDL of 21~m WiSEs, and other concentrated electrolytes, remains to be fully understood~\cite{Zhou2022Agg}. Our predictions, however, appear to match experimental findings remarkably well. Specifically, at negative electrodes, we predicted two distinct layers of Li$^+$. Right at the interface, we predicted it to be dominated by hydrated cations, while the second layer contains some aggregates, before the bulk gel phase is reached. This is remarkably similar to the AFM and SFA measurements of Han \textit{et al.}~\cite{Han2021WiSE}, where the first layer was found to be composed of hydrated Li$^+$, the second layer containing some ion pairs, before the aggregate size would sharply increase, with nanostructure being detected as far out as 60~nm. 

\section{Conclusion}

Overall, we extended our comparison between atomistic molecular dynamics simulations and theory of the EDL in LiTFSI WiSE to 21~m. To achieve this from theory, we used the Boltzmann closure relations and introduce a charge rescaling factor. Qualitatively, good agreement with theory and simulation was found. This theory was then used to predict thermodynamic effects which could influence the stability and reactivity towards interphase formation. We first computed the changes in activity in the bulk, where we found these changes should result in a positive increase in the redox potentials for the formation of Li salts, such as LiF, in the SEI and thermodynamically suppress the hydrogen evolution reaction at negative potentials. The influence of the activity change on the oxygen evolution reaction are also explored. In addition, we used the equilibrium EDL concentrations to make predictions about how the kinetics of these reactions could be affected, where we found the formation of aggregates in the EDL should facilitate the formation of Li salts in the SEI. We presented a discussion of how this formalism could be extended to more complicated electrolytes, and applied to a wide variety of energy storage technologies. 

\section{Acknowledgments}

We are grateful to J. Pedro de Souza for the helpful discussions. D.M.M. \& M.Z.B. acknowledge support from the Center for Enhanced Nanofluidic Transport 2 (CENT$^2$), an Energy Frontier Research Center funded by the U.S. Department of Energy (DOE), Office of Science, Basic Energy Sciences (BES), under award \# DE-SC0019112. D.M.M. also acknowledges support from the National Science Foundation Graduate Research Fellowship under Grant No. 2141064. M.M. and M.Z.B. acknowledge support from an Amar G. Bose Research Grant. Z.A.H.G acknowledges support through the Glasstone Research Fellowship in Materials and The Queen's College, University of Oxford. Q.Z. \& R.M.E.-M thanks the National Science Foundation for partially funding this research under National Science Foundation grants DMR 1904681, CBET 1916609 and CBET 2516268. This work was partially supported by the U.S. Army DEVCOM ARL Army Research Office (ARO) Energy Sciences Competency, Electrochemistry Program award \# W911NF-24-1-0209. The views and conclusions contained in this document are those of the authors and should not be interpreted as representing the official policies, either expressed or implied, of the U.S. Army or the U.S. Government. The authors acknowledge funding from the Faraday Institution (faraday.ac.uk; EP/S003053/1 grant numbers FIRG060, and FIRG082), the European Research Council (ERC) under the European Union’s Horizon 2020 research and innovation programme (EXISTAR, grant agreement No. 950598) and a UKRI Future Leaders Fellowship (MR/V024558/1).

\section{Data Availability}

Data and scripts for this paper are available at \url{https://github.com/DMMarkiewitz/Faraday_Discussions_2026}. This includes scripts to solve the Boltzmann closure relation, implicit Poisson solver for EDL quantities, MD analysis scripts, MD data, and scripts to produce the figures comparing simulations and theory.

\section{Conflicts of Interest}

The authors declare no conflicts of interest.

\bibliography{WiSE}

@article{Cross2026,
  title={Short-range electrostatic screening in ionic liquids as inferred
by direct force measurements},
  author={Benjamin Cross and Leo Garcia and Elisabeth Charlaix and Patrick Kekicheff},
  journal={PNAS},
  volume={123},
  pages={e2517939123},
  year={2026},
}

@article{Wen2023net,
  title={Chemical reaction networks and opportunities for machine learning},
  author={Mingjian Wen and Evan Walter Clark Spotte-Smith and Samuel M. Blau and Matthew J. McDermott and Aditi S. Krishnapriyan and Kristin A. Persson},
  journal={Nature Computational Science},
  volume={3},
  pages={12--24},
  year={2023},
}

@article{Abraham2025,
  title={Ab Initio Free Energy Surfaces for Coupled Ion-Electron Transfer},
  author={Ethan Abraham and Martin Z Bazant and Troy Van Voorhis},
  journal={arXiv:2510.19106},
  volume={},
  pages={},
  year={2025},
}

@article{Swallow2022Reveal,
  title={Revealing solid electrolyte interphase formation through interface-sensitive Operando X-ray absorption spectroscopy},
  author={Jack E. N. Swallow and Michael W. Fraser and Nis-Julian H. Kneusels and Jodie F. Charlton and Christopher G. Sole and Conor M. E. Phelan and Erik Bj\"orklund and Peter Bencok and Carlos Escudero and Virginia P\'erez-Dieste and Clare P. Grey and Rebecca J. Nicholls and Robert S. Weatherup },
  journal={Nat. Commun.},
  volume={13},
  pages={6070},
  year={2022},
}

@article{Gebbie2023,
  title={Linking Electric Double Layer Formation to Electrocatalytic Activity},
  author={Matthew A. Gebbie and Beichen Liu and Wenxiao Guo and Seth R. Anderson and Samuel G. Johnstone},
  journal={ACS Catal.},
  volume={13},
  pages={16222--16239},
  year={2023},
}

@article{Qisheng23JACS,
  title={Effect of the Electric Double Layer (EDL) in Multicomponent Electrolyte Reduction and Solid Electrolyte Interphase (SEI) Formation in Lithium Batteries},
  author={Qisheng Wu and Matthew T. McDowell and Yue Qi},
  journal={JACS},
  volume={145},
  pages={2473--2484},
  year={2023},
}

@article{Juraskova2025ML,
  title={Modelling ligand exchange in metal complexes with machine learning potentials},
  author={Veronika Juraskova and Gers Tusha and Hanwen Zhang and Lars V Sch\"afer and Fernanda Duarte},
  journal={Faraday Discussion},
  volume={256},
  pages={156-176},
  year={2025},
}

@article{Zhang24ML,
  title={Modelling chemical processes in explicit solvents with machine learning potentials},
  author={Hanwen Zhang and Veronika Juraskova and Fernanda Duarte},
  journal={Nat. Commun},
  volume={15},
  pages={6114},
  year={2024},
}

@article{Yang2025ML,
  title={Room-temperature decomposition of the ethaline deep eutectic solvent},
  author={Julia H Yang and Amanda Whai Shin Ooi and Zachary AH Goodwin and Yu Xie and Jingxuan Ding and Stefano Falletta and Ah-Hyung Alissa Park and Boris Kozinsky},
  journal={J. Phys. Chem. Lett.},
  volume={16},
  pages={3039--3046},
  year={2025},
}

@article{Goodwin2024ML,
  title={Transferability and accuracy of ionic liquid simulations with equivariant machine learning interatomic potentials},
  author={Zachary AH Goodwin and Malia B Wenny and Julia H Yang and Andrea Cepellotti and Jingxuan Ding and Kyle Bystrom and Blake R Duschatko and Anders Johansson and Lixin Sun and Simon Batzner and Albert Musaelian and Jarad A Mason and Boris Kozinsky and Nicola Molinari},
  journal={J. Phys. Chem. Lett.},
  volume={15},
  pages={7539--7547},
  year={2024},
}

@article{takenaka2024liquid,
  title={Liquid Madelung energy accounts for the huge potential shift in electrochemical systems},
  author={Takenaka, Norio and Ko, Seongjae and Kitada, Atsushi and Yamada, Atsuo},
  journal={Nature Communications},
  volume={15},
  number={1},
  pages={1319},
  year={2024},
  publisher={Nature Publishing Group UK London}
}

@article{Cheol2023HE,
  title={High-entropy electrolytes for practical lithium metal batteries},
  author={Sang Cheol Kim and Jingyang Wang and Rong Xu and Pu Zhang and Yuelang Chen and Zhuojun Huang and Yufei Yang and Zhiao Yu and Solomon T. Oyakhire and Wenbo Zhang and Louisa C. Greenburg and Mun Sek Kim and David T. Boyle and Philaphon Sayavong and Yusheng Ye and Jian Qin and Zhenan Bao and Yi Cui },
  journal={Nature Energy},
  volume={8},
  pages={814--826},
  year={(2023)},
}

@phdthesis{McEldrewPhD,
    author = {Michael McEldrew},
    title = {Ion Aggregation, Correlated Ion Transport and the Double Layer in Super-Concentrated Electrolytes},
    school = {MIT},
    year = {2021}
}

@article{Tkachenko2025,
  title={Dehydration-Driven Ion Aggregation and the Onset of Gelation in ZnCl2 Solution},
  author={Alexei V. Tkachenko and Chuntian Cao and Amy C. Marschilok and Deyu Lu},
  journal={arXiv:2512.18167},
  volume={},
  pages={},
  year={2025},
}

@article{Finney2021,
  title={Electrochemistry, ion adsorption and dynamics in the double layer: a study of NaCl(aq) on graphite},
  author={Aaron R. Finney and Ian J. McPherson and Patrick R. Unwin and Matteo Salvalaglio},
  journal={Chemical Science},
  volume={12},
  pages={11166},
  year={2021},
}

@article{GoodwinHelm2025,
  title={Theory of Cation Solvation in the Helmholtz Layer of Li-Ion Battery Electrolytes},
  author={Zachary AH Goodwin and Daniel M Markiewitz and Qisheng Wu and Yue Qi and Martin Z Bazant},
  journal={ACS Applied Energy Materials},
  volume={8},
  pages={8376--8387},
  year={2025},
}

@article{Daghetti1993,
  title={Single-ion Activities Based on the Electrical Double-layer Model: An Indirect Test of the Gouy-Chapman Theory},
  author={Anna Daghetti and Simona Romeo and Maurizio Usuelli and Sergio Trasatti},
  journal={J. CHEM. SOC. FARADAY TRANS.},
  volume={89},
  pages={187-193},
  year={1993},
}

@article{Markiewitz2025,
  title={Ionic associations and hydration in the electrical double layer of water-in-salt electrolytes},
  author={Daniel M Markiewitz and Zachary AH Goodwin and Qianlu Zheng and Michael McEldrew and Rosa M Espinosa-Marzal and Martin Z Bazant},
  journal={ACS Applied Materials \& Interfaces},
  volume={17},
  pages={29515-29534},
  year={2025},
}

@article{Han2021WiSE,
  title={Nanoheterogeneity of LiTFSI Solutions Transitions Close to a Surface and with Concentration},
  author={Mengwei Han and Ruixian Zhang and Andrew A. Gewirth and Rosa M. Espinosa-Marzal},
  journal={Nano Lett.},
  volume={21},
  pages={2304--2309},
  year={2021},
}

@article{Goodwin2021Rev,
  title={Mean-Field Theory of the Electrical Double Layer in Ionic Liquids},
  author={Zachary A. H. Goodwin and J. Pedro de Souza and Martin Z. Bazant and Alexei A. Kornyshev},
  journal={Encyclopedia of Ionic Liquids},
  volume={},
  pages={1--13},
  year={2021},
}

@article{Pedro2020,
  title={Interfacial layering in the electrical double layer of ionic liquids},
  author={J Pedro de Souza and Zachary AH Goodwin and Michael McEldrew and Alexei A Kornyshev and Martin Z Bazant},
  journal={Phys. Rev. Lett.},
  volume={125},
  pages={116001},
  year={2020},
}

@article{Fraggedakis2021,
  title={Theory of coupled ion-electron transfer kinetics},
  author={Dimitrios Fraggedakis and Michael McEldrew and Raymond B. Smith and Yamini Krishnan and Yirui Zhang and Peng Bai and William C. Chueh and Yang Shao-Horn and Martin Z.Bazant},
  journal={Electrochim. Acta},
  volume={367},
  pages={137432},
  year={2021},
}

@article{Han2020,
  title={Insight into the Electrical Double Layer of Ionic Liquids Revealed
through Its Temporal Evolution},
  author={M Han and H Kim and C Leal and M Negrito and J D Batteas and R M Espinosa-Marzal},
  journal={Adv Mater Interfaces},
  volume={7},
  pages={2001313},
  year={2020},
}

@inproceedings{Espinosa2023rev,
  title={Colloidal Interactions in Ionic Liquids—The Electrical Double Layer Inferred from Ion Layering and Aggregation},
  author={Espinosa-Marzal, Rosa M and Goodwin, Zachary AH and Zhang, Xuhui and Zheng, Qianlu},
  booktitle={One Hundred Years of Colloid Symposia: Looking Back and Looking Forward},
  pages={123--148},
  year={2023},
  organization={ACS Publications}
}

@article{Zhou2022Agg,
  title={Beyond Local Solvation Structure: Nanometric Aggregates in Battery Electrolytes and Their Effect on Electrolyte Properties},
  author={Zhou Yu and Nitash P. Balsara and Oleg Borodin and Andrew A. Gewirth and Nathan T. Hahn and Edward J. Maginn and Kristin A. Persson and Venkat Srinivasan and Michael F. Toney and Kang Xu and Kevin R. Zavadil and Larry A. Curtiss and Lei Cheng},
  journal={ACS Energy Lett.},
  volume={7},
  pages={461--470},
  year={2022},
}

@article{Wang2023HEE2,
  title={High entropy liquid electrolytes for lithium batteries},
  author={Qidi Wang and Chenglong Zhao and Jianlin Wang and Zhenpeng Yao and Shuwei Wang and Sai Govind Hari Kumar and Swapna Ganapathy and Stephen Eustace and Xuedong Bai and Baohua Li and Marnix Wagemaker},
  journal={Nat. Commun.},
  volume={14},
  pages={440},
  year={2023},
}

@article{Goodwin2022Kornyshev,
  title={Cracking Ion Pairs in the Electrical Double Layer of Ionic Liquids},
  author={Zachary AH Goodwin and Alexei A Kornyshev},
  journal={Electrochim. Acta},
  volume={434},
  pages={141163},
  year={2022},
}

@article{Goodwin2022EDL,
  title={Gelation, Clustering and Crowding in the Electrical Double Layer of Ionic Liquids},
  author={Zachary AH Goodwin and Michael McEldrew and J Pedro de Souza and Martin Z Bazant and Alexei A Kornyshev},
  journal={J. Chem. Phys.},
  volume={157},
  pages={094106},
  year={2022},
}

@article{Goodwin2023,
  title={Theory of Cation Solvation and Ionic Association in Nonaqueous Solvent Mixtures},
  author={Zachary A H Goodwin and Michael McEldrew and Boris Kozinsky and Martin Z Bazant},
  journal={PRX Energy},
  volume={2},
  pages={013007},
  year={2023},
}

@article{McEldrewsalt2021,
  title={Salt-in-ionic-liquid electrolytes: Ion network formation and negative effective charges of alkali metal cations},
  author={Michael McEldrew and Zachary A H Goodwin and Nicola Molinari and Boris Kozinsky and Alexei A Kornyshev and Martin Z Bazant},
  journal={J. Phys. Chem. B},
  volume={125},
  pages={13752--13766},
  year={2021},
}

@article{mceldrew2021ion,
  title={Ion Clusters and Networks in Water-in-Salt Electrolytes},
  author={McEldrew, Michael and Goodwin, Zachary AH and Bi, Sheng and Kornyshev, Alexei and Bazant, Martin Z},
  journal={J. Electrochem. Soc.},
  year={2021},
  volume = {168},
  pages = {050514},
  publisher={IOP Publishing}
}

@article{mceldrew2020corr,
  title={Correlated Ion Transport and the Gel Phase in Room Temperature Ionic Liquids},
  author={Michael McEldrew and Zachary A. H. Goodwin and Hongbo Zhao and Martin Z. Bazant and Alexei A. Kornyshev},
  journal={J. Phys. Chem B},
  volume={125},
  pages={2677–2689},
  year={2021},
}

@article{kilic2007a,
  title={Steric effects in the dynamics of electrolytes at large applied voltages. I. Double-layer charging},
  author={Kilic, Mustafa Sabri and Bazant, Martin Z and Ajdari, Armand},
  journal={Physical review E},
  volume={75},
  number={2},
  pages={021502},
  year={2007},
  publisher={APS}
}

@article{Bazant2009a,
    title={Towards an understanding of induced-charge electrokinetics at large applied voltages in concentrated solutions},
    author={Bazant, Martin Z and Kilic, Mustafa Sabri and Storey, Brian D and Ajdari, Armand},
    journal={Advances in colloid and interface science},
    volume={152},
    number={1-2},
    pages={48--88},
    year={2009},
    publisher={Elsevier}
}

@article{Gebbie2013,
author = {Gebbie, M. A. and Valtiner, M. and Banquy, X. and Fox, E. T. and Henderson, W. A. and Israelachvili, J. N.},
doi = {10.1073/pnas.1307871110},
issn = {0027-8424},
journal = {Proceedings of the National Academy of Sciences},
month = {jun},
number = {24},
pages = {9674--9679},
title = {{Ionic liquids behave as dilute electrolyte solutions}},
url = {http://www.pnas.org/cgi/doi/10.1073/pnas.1307871110},
volume = {110},
year = {2013}
}

@article{Gebbie2015,
author = {Gebbie, M. A. and Dobes, H. A. and Valtiner, M. and Israelachvili, J. N.},
journal = {Proceedings of the National Academy of Sciences},
pages = {7432–7437},
title = {{Long-range electrostatic screening in ionic liquids}},
volume = {112},
year = {2015}
}

@article{Goodwin2017a,
author = {Goodwin, Zachary A. H. and Feng, Guang and Kornyshev, Alexei A.},
doi = {10.1016/J.ELECTACTA.2016.12.092},
issn = {0013-4686},
journal = {Electrochim. Acta},
month = {jan},
pages = {190--197},
publisher = {Pergamon},
title = {{Mean-Field Theory of Electrical Double Layer In Ionic Liquids with Account of Short-Range Correlations}},
url = {https://www.sciencedirect.com/science/article/pii/S001346861632638X},
volume = {225},
year = {2017}
}

@article{vatamanu2017,
  title={Ramifications of water-in-salt interfacial structure at charged electrodes for electrolyte electrochemical stability},
  author={Vatamanu, Jenel and Borodin, Oleg},
  journal={J. Phys. Chem. Lett.},
  volume={8},
  number={18},
  pages={4362--4367},
  year={2017},
  publisher={ACS Publications}
}

@article{Fedorov2014,
author = {Fedorov, Maxim V. and Kornyshev, Alexei A.},
doi = {10.1021/cr400374x},
issn = {0009-2665},
journal = {Chem. Rev.},
month = {mar},
number = {5},
pages = {2978--3036},
title = {{Ionic Liquids at Electrified Interfaces}},
url = {http://pubs.acs.org/doi/10.1021/cr400374x},
volume = {114},
year = {2014}
}

@article{Kornyshev2007,
author = {Kornyshev, Alexei A.},
doi = {10.1021/jp067857o},
issn = {1520-6106},
journal = {J. Phys. Chem. B},
month = {may},
number = {20},
pages = {5545--5557},
title = {{Double-Layer in Ionic Liquids:  Paradigm Change?}},
url = {http://pubs.acs.org/doi/abs/10.1021/jp067857o},
volume = {111},
year = {2007}
}

@article{Suo2015,
author = {Suo, Liumin and Borodin, Oleg and Gao, Tao and Olguin, Marco and Ho, Janet and Fan, Xiulin and Luo, Chao and Wang, Chunsheng and Xu, Kang},
doi = {10.1126/science.aab1595},
issn = {1095-9203},
journal = {Science},
month = {nov},
number = {6263},
pages = {938-43},
pmid = {26586759},
publisher = {American Association for the Advancement of Science},
title = {{``Water-in-salt" electrolyte enables high-voltage aqueous lithium-ion chemistries.}},
url = {http://www.ncbi.nlm.nih.gov/pubmed/26586759},
volume = {350},
year = {2015}
}

@article{Suo2013,
author = {Suo, Liumin and Hu, Yong-Sheng and Li, Hong and Armand, Michel and Chen, Liquan},
doi = {10.1038/ncomms2513},
issn = {2041-1723},
journal = {Nat. Commun.},
month = {dec},
number = {1},
pages = {1481},
publisher = {Nature Publishing Group},
title = {{A new class of Solvent-in-Salt electrolyte for high-energy rechargeable metallic lithium batteries}},
url = {http://www.nature.com/articles/ncomms2513},
volume = {4},
year = {2013}
}

@article{Wang2016,
author = {Wang, Jianhui and Yamada, Yuki and Sodeyama, Keitaro and Chiang, Ching Hua and Tateyama, Yoshitaka and Yamada, Atsuo},
doi = {10.1038/ncomms12032},
issn = {2041-1723},
journal = {Nat. Commun.},
month = {jun},
pages = {12032},
publisher = {Nature Publishing Group},
title = {{Superconcentrated electrolytes for a high-voltage lithium-ion battery}},
url = {http://www.nature.com/doifinder/10.1038/ncomms12032},
volume = {7},
year = {2016}
}

@article{Suo2016,
author = {Suo, Liumin and Borodin, Oleg and Sun, Wei and Fan, Xiulin and Yang, Chongyin and Wang, Fei and Gao, Tao and Ma, Zhaohui and Schroeder, Marshall and von Cresce, Arthur and Russell, Selena M. and Armand, Michel and Angell, Austen and Xu, Kang and Wang, Chunsheng},
doi = {10.1002/anie.201602397},
issn = {14337851},
journal = {Angew. Chem.},
month = {jun},
number = {25},
pages = {7136--7141},
publisher = {Wiley-Blackwell},
title = {{Advanced High-Voltage Aqueous Lithium-Ion Battery Enabled by “Water-in-Bisalt” Electrolyte}},
url = {http://doi.wiley.com/10.1002/anie.201602397},
volume = {55},
year = {2016}
}

@article{Yamada2016,
author = {Yamada, Yuki and Usui, Kenji and Sodeyama, Keitaro and Ko, Seongjae and Tateyama, Yoshitaka and Yamada, Atsuo},
doi = {10.1038/nenergy.2016.129},
issn = {2058-7546},
journal = {Nat. Energy},
month = {aug},
number = {10},
pages = {16129},
publisher = {Nature Publishing Group},
title = {{Hydrate-melt electrolytes for high-energy-density aqueous batteries}},
url = {http://www.nature.com/articles/nenergy2016129},
volume = {1},
year = {2016}
}

@article{xu2004nonaqueous,
  title={Nonaqueous liquid electrolytes for lithium-based rechargeable batteries},
  author={Xu, Kang},
  journal={Chem. Rev.},
  volume={104},
  number={10},
  pages={4303--4418},
  year={2004},
  publisher={ACS Publications}
}

@article{borodin2017liquid,
  title={Liquid structure with nano-heterogeneity promotes cationic transport in concentrated electrolytes},
  author={Borodin, Oleg and Suo, Liumin and Gobet, Mallory and Ren, Xiaoming and Wang, Fei and Faraone, Antonio and Peng, Jing and Olguin, Marco and Schroeder, Marshall and Ding, Michael S and others},
  journal={ACS nano},
  volume={11},
  number={10},
  pages={10462--10471},
  year={2017},
  publisher={ACS Publications}
}

@article{wang2018hybrid,
  title={Hybrid aqueous/non-aqueous electrolyte for safe and high-energy Li-ion batteries},
  author={Wang, Fei and Borodin, Oleg and Ding, Michael S and Gobet, Mallory and Vatamanu, Jenel and Fan, Xiulin and Gao, Tao and Edison, Nico and Liang, Yujia and Sun, Wei and others},
  journal={Joule},
  volume={2},
  number={5},
  pages={927--937},
  year={2018},
  publisher={Elsevier}
}

@article{Borodin2017Mod,
  title={Modeling Insight into Battery Electrolyte Electrochemical Stability and Interfacial Structure},
  author={Oleg Borodin and Xiaoming Ren and Jenel Vatamanu and Arthur von Wald Cresce and Jaroslaw Knap and Kang Xu},
  journal={Acc. Chem. Res.},
  volume={50},
  pages={2886--2894},
  year={2017},
}

@article{Ravikumar2018,
  title={Effect of Salt Concentration on Properties of Lithium Ion Battery Electrolytes: A Molecular Dynamics Study},
  author={Bharath Ravikumar and Mahesh Mynam and Beena Rai},
  journal={J. Phys. Chem. C},
  volume={122},
  pages={8173--8181},
  year={2018},
}

@article{Shim2018,
  title={Computer simulation study of the solvation of lithium ions in ternary mixed carbonate electrolytes: free energetics, dynamics, and ion transport},
  author={Youngseon Shim},
  journal={Phys. Chem. Chem. Phys.},
  volume={20},
  pages={28649},
  year={2018},
}

@article{Han2019MD,
  title={Structure and dynamics in the lithium solvation shell of nonaqueous electrolytes},
  author={Sungho Han},
  journal={Sci. Rep.},
  volume={9},
  pages={5555},
  year={2019},
}

@article{Piao2020Count,
  title={Countersolvent Electrolytes for Lithium-Metal Batteries},
  author={Nan Piao and Xiao Ji and Hong Xu and Xiulin Fan and Long Chen and Sufu Liu and Mounesha N. Garaga and Steven G. Greenbaum and Li Wang and Chunsheng Wang and Xiangming He},
  journal={Adv. Energy Mater.},
  volume={10},
  pages={1903568},
  year={2020},
}

@article{Hou2021,
  title={The solvation structure, transport properties and reduction behavior of carbonate-based electrolytes of lithium-ion batteries},
  author={Tingzheng Hou and Kara D. Fong and Jingyang Wang and Kristin A. Persson},
  journal={Chem. Sci.},
  volume={12},
  pages={14740},
  year={2021},
}

@article{Han2017MD,
  title={A salient effect of density on the dynamics of nonaqueous electrolytes},
  author={Sungho Han},
  journal={Sci. Rep.},
  volume={7},
  pages={46718},
  year={2017},
}

@article{lim2018,
  title={Nanometric water channels in water-in-salt lithium ion battery electrolyte},
  author={Lim, Joonhyung and Park, Kwanghee and Lee, Hochan and Kim, Jungyu and Kwak, Kyungwon and Cho, Minhaeng},
  journal={Journal of the American Chemical Society},
  volume={140},
  number={46},
  pages={15661--15667},
  year={2018},
  publisher={ACS Publications}
}

@article{mceldrew2018,
  title={Theory of the double layer in water-in-salt electrolytes},
  author={McEldrew, Michael and Goodwin, Zachary AH and Kornyshev, Alexei A and Bazant, Martin Z},
  journal={The journal of physical chemistry letters},
  volume={9},
  number={19},
  pages={5840--5846},
  year={2018},
  publisher={ACS Publications}
}

@article{molinari2019transport,
  title={Transport anomalies emerging from strong correlation in ionic liquid electrolytes},
  author={Molinari, Nicola and Mailoa, Jonathan P and Craig, Nathan and Christensen, Jake and Kozinsky, Boris},
  journal={J. Power Sources},
  volume={428},
  pages={27--36},
  year={2019},
  publisher={Elsevier}
}

@article{molinari2019general,
  title={General Trend of a Negative Li Effective Charge in Ionic Liquid Electrolytes},
  author={Molinari, Nicola and Mailoa, Jonathan P and Kozinsky, Boris},
  journal={J. Phys. Chem. Lett.},
  volume={10},
  number={10},
  pages={2313--2319},
  year={2019},
  publisher={ACS Publications}
}

@article{borodin2020uncharted,
  title={Uncharted Waters: Super-Concentrated Electrolytes},
  author={Borodin, Oleg and Self, Julian and Persson, Kristin A and Wang, Chunsheng and Xu, Kang},
  journal={Joule},
  volume={4},
  number={1},
  pages={69--100},
  year={2020},
  publisher={Elsevier}
}

@article{suo2017water,
  title={“Water-in-salt” electrolyte makes aqueous sodium-ion battery safe, green, and long-lasting},
  author={Suo, Liumin and Borodin, Oleg and Wang, Yuesheng and Rong, Xiaohui and Sun, Wei and Fan, Xiiulin and Xu, Shuyin and Schroeder, Marshall A and Cresce, Arthur V and Wang, Fei and others},
  journal={Advanced Energy Materials},
  volume={7},
  number={21},
  pages={1701189},
  year={2017},
  publisher={Wiley Online Library}
}

@article{Wang2018,
author = {Wang, Fei and Borodin, Oleg and Gao, Tao and Fan, Xiulin and Sun, Wei and Han, Fudong and Faraone, Antonio and Dura, Joseph A. and Xu, Kang and Wang, Chunsheng},
doi = {10.1038/s41563-018-0063-z},
issn = {14764660},
journal = {Nature Materials},
number = {6},
pages={543--549},
title = {{Highly reversible zinc metal anode for aqueous batteries}},
volume = {17},
year = {2018}
}

@article{Dou2019,
author = {Dou, Qingyun and Lu, Yulan and Su, Lijun and Zhang, Xu and Lei, Shulai and Bu, Xudong and Liu, Lingyang and Xiao, Dewei and Chen, Jiangtao and Shi, Siqi and Yan, Xingbin},
doi = {10.1016/j.ensm.2019.03.016},
issn = {24058297},
journal = {Energy Storage Materials},
title = {{A sodium perchlorate-based hybrid electrolyte with high salt-to-water molar ratio for safe 2.5 V carbon-based supercapacitor}},
volume = {23},
year = {2019}
}

@article{smith2016electrostatic,
  title={The electrostatic screening length in concentrated electrolytes increases with concentration},
  author={Smith, Alexander M and Lee, Alpha A and Perkin, Susan},
  journal={J. Phys. Chem. Lett.},
  volume={7},
  number={12},
  pages={2157--2163},
  year={2016},
  publisher={ACS Publications}
}

@article{mceldrew2020theory,
  title={Theory of Ion Aggregation and Gelation in Super-Concentrated Electrolytes},
  author={McEldrew, Michael and Goodwin, Zachary AH and Bi, Sheng and Bazant, Martin Z and Kornyshev, Alexei A},
  journal={J. Chem. Phys.},
  volume={152},
  pages={234506},
  year={2020}
}

@article{chen202063,
  title={A 63 m Super-concentrated Aqueous Electrolyte for High Energy Li-ion Batteries},
  author={Chen, Long and Zhang, Jiaxun and Li, Qin and Vatamanu, Jenel and Ji, Xiao and Pollard, Travis P and Cui, Chunyu and Hou, Singyuk and Chen, Ji and Yang, Chongyin and others},
  journal={ACS Energy Lett.},
  year={2020},
  publisher={ACS Publications}
}

@article{li2020new,
  title={New Concepts in Electrolytes},
  author={Li, Matthew and Wang, Chunsheng and Chen, Zhongwei and Xu, Kang and Lu, Jun},
  journal={Chem. Rev.},
  year={2020},
  publisher={ACS Publications}
}

@article{Kang2022Nav,
  title={Navigating the minefield of battery literature},
  author={Kang Xu},
  journal={Commun. Mater},
  volume={3},
  pages={31},
  year={2022},
}

@article{Cheng2022Sol,
  title={Emerging Era of Electrolyte Solvation
Structure and Interfacial Model in Batteries},
  author={Haoran Cheng and Qujiang Sun and Leilei Li and Yeguo Zou and Yuqi Wang and Tao Cai and Fei Zhao and Gang Liu and Zheng Ma and Wandi Wahyudi and Qian Li and Jun Ming},
  journal={ACS Energy Lett.},
  volume={7},
  pages={490--513},
  year={2022},
}

@article{Xie2023,
  title={Spatially resolved structural order in low-temperature liquid electrolyte},
  author={Yujun Xie and Jingyang Wang and Benjamin H. Savitzky and Zheng Chen and Yu Wang and Sophia Betzler and Karen Bustillo and Kristin Persson and Yi Cui and Lin-Wang Wang and Colin Ophus and Peter Ercius and Haimei Zheng},
  journal={Sci. Adv.},
  volume={9},
  pages={eadc9721},
  year={2023},
}

@article{Kim2021Pot,
  title={Potentiometric Measurement to Probe Solvation Energy and Its Correlation to Lithium Battery Cyclability},
  author={Sang Cheol Kim and Xian Kong and Rafael A. Vil\'a and William Huang and Yuelang Chen and David T. Boyle and Zhiao Yu and Hansen Wang and Zhenan Bao and Jian Qin and Yi Cui},
  journal={J. Am. Chem. Soc.},
  volume={143},
  pages={10301--10308},
  year={2021},
}

@article{Li2015,
  title={Effect of Organic Solvents on {L}i$^+$ Ion Solvation and Transport in Ionic Liquid Electrolytes: A Molecular Dynamics Simulation Study},
  author={Zhe Li and Oleg Borodin and Grant D. Smith and Dmitry Bedrov},
  journal={J. Phys. Chem. B},
  volume={119},
  pages={3085--3096},
  year={2015},
}

@article{Skarmoutsos2015,
  title={Li+ Solvation in Pure, Binary, and Ternary Mixtures of Organic Carbonate Electrolytes},
  author={Ioannis Skarmoutsos and Veerapandian Ponnuchamy and Valentina Vetere and Stefano Mossa},
  journal={J. Phys. Chem. C},
  volume={119},
  pages={4502--4515},
  year={2015},
}

@article{Postupna2011,
  title={Microscopic Structure and Dynamics of LiBF4 Solutions in Cyclic and Linear Carbonates},
  author={O. O. Postupna and Y. V. Kolesnik and O. N. Kalugin and O. V. Prezhdo},
  journal={J. Phys. Chem. B},
  volume={115},
  pages={14563--14571},
  year={2011},
}

@article{Borodin2014SEI,
  title={Interfacial Structure and Dynamics of the Lithium Alkyl Dicarbonate SEI Components in Contact with the Lithium Battery Electrolyte},
  author={Oleg Borodin and Dmitry Bedrov},
  journal={J. Phys. Chem. C},
  volume={118},
  pages={18362--18371},
  year={2014},
}

@article{wang2022liquid,
  title={Liquid electrolyte: The nexus of practical lithium metal batteries},
  author={Wang, Hansen and Yu, Zhiao and Kong, Xian and Kim, Sang Cheol and Boyle, David T and Qin, Jian and Bao, Zhenan and Cui, Yi},
  journal={Joule},
  year={2022},
  volume = {6},
  pages = {588--616},
  publisher={Elsevier}
}

@article{Wang2021corsol,
  title={Correlating Li-Ion Solvation Structures and Electrode Potential Temperature Coefficients},
  author={Hansen Wang and Sang Cheol Kim and Toma\'as Rojas and Yangying Zhu and Yanbin {L}i and Lin Ma and Kang Xu and Anh T. Ngo and Yi Cui},
  journal={J. Am. Chem. Soc.},
  volume={143},
  pages={2264--2271},
  year={2021},
}

@article{yu2020molecular,
  title={Molecular design for electrolyte solvents enabling energy-dense and long-cycling lithium metal batteries},
  author={Yu, Zhiao and Wang, Hansen and Kong, Xian and Huang, William and Tsao, Yuchi and Mackanic, David G and Wang, Kecheng and Wang, Xinchang and Huang, Wenxiao and Choudhury, Snehashis and others},
  journal={Nat. Energy},
  volume={5},
  number={7},
  pages={526--533},
  year={2020},
  publisher={Nature Publishing Group}
}

@article{Piao2022,
  title={A review on regulating {L}i$^+$ solvation structures in carbonate electrolytes for lithium metal batteries},
  author={Zhihong Piao and Runhua Gao and Yingqi Liu and Guangmin Zhou and Hui-Ming Cheng},
  journal={Adv.Mater.},
  volume={},
  pages={2206009},
  year={2023},
}

@article{yu2022rational,
  title={Rational solvent molecule tuning for high-performance lithium metal battery electrolytes},
  author={Yu, Zhiao and Rudnicki, Paul E and Zhang, Zewen and Huang, Zhuojun and Celik, Hasan and Oyakhire, Solomon T and Chen, Yuelang and Kong, Xian and Kim, Sang Cheol and Xiao, Xin and others},
  journal={Nat. Energy},
  volume={7},
  number={1},
  pages={94--106},
  year={2022},
  publisher={Nature Publishing Group}
}

@article{qin2019localized,
  title={Localized high-concentration electrolytes boost potassium storage in high-loading graphite},
  author={Qin, Lei and Xiao, Neng and Zheng, Jingfeng and Lei, Yu and Zhai, Dengyun and Wu, Yiying},
  journal={Adv. Energy Mater.},
  volume={9},
  number={44},
  pages={1902618},
  year={2019},
  publisher={Wiley Online Library}
}

@article{zheng2018extremely,
  title={Extremely stable sodium metal batteries enabled by localized high-concentration electrolytes},
  author={Zheng, Jianming and Chen, Shuru and Zhao, Wengao and Song, Junhua and Engelhard, Mark H and Zhang, Ji-Guang},
  journal={ACS Energy Lett.},
  volume={3},
  number={2},
  pages={315--321},
  year={2018},
  publisher={ACS Publications}
}

@article{chen2021highly,
  title={Highly reversible aqueous zinc metal batteries enabled by fluorinated interphases in localized high concentration electrolytes},
  author={Chen, Shunqiang and Nian, Qingshun and Zheng, Lei and Xiong, Bing-Qing and Wang, Zihong and Shen, Yanbin and Ren, Xiaodi},
  journal={J. Mater. Chem. A},
  volume={9},
  number={39},
  pages={22347--22352},
  year={2021},
  publisher={Royal Society of Chemistry}
}

@article{li2020high,
  title={High-nickel layered oxide cathodes for lithium-based automotive batteries},
  author={Li, Wangda and Erickson, Evan M and Manthiram, Arumugam},
  journal={Nat. Energy},
  volume={5},
  number={1},
  pages={26--34},
  year={2020},
  publisher={Nature Publishing Group}
}

@article{Tian2021,
  title={Promises and Challenges of Next-Generation ``Beyond {L}i-ion'' Batteries for Electric Vehicles and Grid Decarbonization},
  author={Yaosen Tian and Guobo Zeng and Ann Rutt and Tan Shi and Haegyeom Kim and Jingyang Wang and Julius Koettgen and Yingzhi Sun and Bin Ouyang and Tina Chen and Zhengyan Lun and Ziqin Rong and Kristin Persson and Gerbrand Ceder},
  journal={Chem. Rev.},
  volume={121},
  pages={1623--1669},
  year={2021},
}

@article{Ioan2022,
  title={Machine Learning Force Field for Molecular Liquids: EC/EMC Binary Solvent},
  author={Ioan-Bogdan Magd\u{a}u and Daniel J. Arismendi-Arrieta and Holly E. Smith and Clare P. Grey and Kersti Hermansson and G\'{a}bor Cs\'{a}nyi},
  journal={npj Computational Materials volume},
  volume={9},
  pages={146},
  year={2023},
}

@article{Yao2022CRev,
  title={Applying Classical, Ab Initio, and Machine-Learning Molecular Dynamics Simulations to the Liquid Electrolyte for Rechargeable Batteries},
  author={Nan Yao and Xiang Chen and Zhong-Heng Fu and Qiang Zhang},
  journal={Chem. Rev.},
  volume={122},
  pages={10970--11021},
  year={2022},
}

@article{zhang2018aqueous,
  title={Aqueous/Nonaqueous Hybrid Electrolyte for Sodium-Ion Batteries},
  author={Zhang, Huang and Qin, Bingsheng and Han, Jin and Passerini, Stefano},
  journal={ACS Energy Lett.},
  volume={3},
  number={7},
  pages={1769--1770},
  year={2018},
  publisher={ACS Publications}
}

@article{xu2007solvation,
  title={Solvation sheath of Li+ in nonaqueous electrolytes and its implication of graphite/electrolyte interface chemistry},
  author={Xu, Kang and Lam, Yiufai and Zhang, Sheng S and Jow, T Richard and Curtis, Timothy B},
  journal={The Journal of Physical Chemistry C},
  volume={111},
  number={20},
  pages={7411--7421},
  year={2007},
  publisher={ACS Publications}
}

@article{von2012correlating,
  title={Correlating Li+ solvation sheath structure with interphasial chemistry on graphite},
  author={von Wald Cresce, Arthur and Borodin, Oleg and Xu, Kang},
  journal={The Journal of Physical Chemistry C},
  volume={116},
  number={50},
  pages={26111--26117},
  year={2012},
  publisher={ACS Publications}
}

@article{yu2020asymmetric,
  title={Asymmetric Composition of Ionic Aggregates and the Origin of High Correlated Transference Number in Water-in-Salt Electrolytes},
  author={Yu, Zhou and Curtiss, Larry A and Winans, Randall E and Zhang, Yang and Li, Tao and Cheng, Lei},
  journal={The Journal of Physical Chemistry Letters},
  volume={11},
  number={4},
  pages={1276--1281},
  year={2020},
  publisher={ACS Publications}
}

@article{zheng2018understanding,
  title={Understanding thermodynamic and kinetic contributions in expanding the stability window of aqueous electrolytes},
  author={Zheng, Jiaxin and Tan, Guoyu and Shan, Peng and Liu, Tongchao and Hu, Jiangtao and Feng, Yancong and Yang, Luyi and Zhang, Mingjian and Chen, Zonghai and Lin, Yuan and others},
  journal={Chem},
  volume={4},
  number={12},
  pages={2872--2882},
  year={2018},
  publisher={Elsevier}
}

@article{lui2011salts,
  title={Salts dissolved in salts: ionic liquid mixtures},
  author={Lui, Matthew Y and Crowhurst, Lorna and Hallett, Jason P and Hunt, Patricia A and Niedermeyer, Heiko and Welton, Tom},
  journal={Chem. Sci.},
  volume={2},
  number={8},
  pages={1491--1496},
  year={2011},
  publisher={Royal Society of Chemistry}
}

@article{jiang2020high,
  title={High-Voltage Aqueous Na-Ion Battery Enabled by Inert-Cation-Assisted Water-in-Salt Electrolyte},
  author={Jiang, Liwei and Liu, Lilu and Yue, Jinming and Zhang, Qiangqiang and Zhou, Anxing and Borodin, Oleg and Suo, Liumin and Li, Hong and Chen, Liquan and Xu, Kang and Hu, Yong‐Sheng},
  journal={ADV MATER},
  volume={32},
  number={2},
  pages={1904427},
  year={2020},
  publisher={Wiley Online Library}
}

@article{molinari2020chelation,
  title={Chelation-Induced Reversal of Negative Cation Transference Number in Ionic Liquid Electrolytes},
  author={Molinari, Nicola and Kozinsky, Boris},
  journal={J. Phys. Chem. B},
  volume={124},
  number={13},
  pages={2676--2684},
  year={2020},
  publisher={ACS Publications}
}

@article{dou2018safe,
  title={Safe and high-rate supercapacitors based on an ``acetonitrile/water in salt” hybrid electrolyte},
  author={Dou, Qingyun and Lei, Shulai and Wang, Da-Wei and Zhang, Qingnuan and Xiao, Dewei and Guo, Hongwei and Wang, Aiping and Yang, Hui and Li, Yongle and Shi, Siqi and Yan, Xingbin},
  journal={Energy Environ. Sci.},
  volume={11},
  number={11},
  pages={3212--3219},
  year={2018},
  publisher={Royal Society of Chemistry}
}

@article{kondou2018enhanced,
  title={Enhanced Electrochemical Stability of Molten Li Salt Hydrate Electrolytes by the Addition of Divalent Cations},
  author={Kondou, Shinji and Nozaki, Erika and Terada, Shoshi and Thomas, Morgan L and Ueno, Kazuhide and Umebayashi, Yasuhiro and Dokko, Kaoru and Watanabe, Masayoshi},
  journal={The Journal of Physical Chemistry C},
  volume={122},
  number={35},
  pages={20167--20175},
  year={2018},
  publisher={ACS Publications}
}

@article{becker2020hybrid,
  title={Hybrid Ionic Liquid/water-in-Salt Electrolytes Enable Stable Cycling of LTO/NMC811 Full Cells},
  author={Becker, Maximilian and Reber, David and Aribia, Abdessalem and Battaglia, Corsin and K{\"u}hnel, Ruben-Simon},
  year={2020},
  publisher={ChemRxiv}
}

@article{steinruck2020interfacial,
  title={Interfacial Speciation Determines Interfacial Chemistry: X-ray-Induced Lithium Fluoride Formation from Water-in-salt Electrolytes on Solid Surfaces},
  author={Steinr{\"u}ck, Hans-Georg and Cao, Chuntian and Lukatskaya, Maria R and Takacs, Christopher J and Wan, Gang and Mackanic, David G and Tsao, Yuchi and Zhao, Jingbo and Helms, Brett A and Xu, Kang and others},
  journal={Angew. Chem.},
  year={2020},
  publisher={Wiley Online Library}
}

@article{andersson2020ion,
  title={Ion Transport Mechanisms via Time-Dependent Local Structure and Dynamics in Highly Concentrated Electrolytes},
  author={Andersson, Rasmus and {\AA}r{\'e}n, Fabian and Franco, Alejandro A and Johansson, Patrik},
  journal={Journal of the Electrochemical Society},
  volume={167},
  number={14},
  pages={140537},
  year={2020},
  publisher={IOP Publishing}
}

@article{Zheng2017Uni,
  title={Research Progress towards Understanding the Unique
Interfaces between Concentrated Electrolytes and
Electrodes for Energy Storage Applications},
  author={Jianming Zheng and Joshua A. Lochala and Alexander Kwok and Zhiqun Daniel Deng and Jie Xiao},
  journal={Adv. Sci.},
  volume={4},
  pages={1700032},
  year={2017},
}

@article{xu2014electrolytes,
  title={Electrolytes and interphases in Li-ion batteries and beyond},
  author={Xu, Kang},
  journal={Chem. Rev.},
  volume={114},
  number={23},
  pages={11503--11618},
  year={2014},
  publisher={ACS Publications}
}

@article{bazant2023unified,
  title={Unified quantum theory of electrochemical kinetics by coupled ion--electron transfer},
  author={Bazant, Martin Z},
  journal={Faraday Discussions},
  volume={246},
  pages={60--124},
  year={2023},
  publisher={Royal Society of Chemistry}
}

@article{Zhang2024,
      title={Long-Range Surface Forces in Salt-in-Ionic Liquids},
      author={Zhang, Xuhui and Goodwin, Zachary AH and Hoane, Alexis G and Deptula, Alex and Markiewitz, Daniel M and Molinari, Nicola and Zheng, Qianlu and Li, Hua and McEldrew, Michael and Kozinsky, Boris and others},
      journal={ACS nano},
      volume={18},
      number={50},
      pages={34007--34022},
      year={2024},
      publisher={ACS Publications}
}

@article{markiewitz2024,
  title={Electric field induced associations in the double layer of salt-in-ionic-liquid electrolytes},
  author={Markiewitz, Daniel M and Goodwin, Zachary AH and McEldrew, Michael and de Souza, J Pedro and Zhang, Xuhui and Espinosa-Marzal, Rosa M and Bazant, Martin Z},
  journal={Faraday Discussions},
  pages={365--384},
  year={2024},
  publisher={Royal Society of Chemistry}
}

@article{sayah2022super,
  title={How do super concentrated electrolytes push the Li-ion batteries and supercapacitors beyond their thermodynamic and electrochemical limits?},
  author={Sayah, Simon and Ghosh, Arunabh and Baazizi, Mariam and Amine, Rachid and Dahbi, Mouad and Amine, Youssef and Ghamouss, Fouad and Amine, Khalil},
  journal={Nano Energy},
  volume={98},
  pages={107336},
  year={2022},
  publisher={Elsevier}
}

@article{zhang2020potential,
  title={Potential-dependent layering in the electrochemical double layer of water-in-salt electrolytes},
  author={Zhang, Ruixian and Han, Mengwei and Ta, Kim and Madsen, Kenneth E and Chen, Xinyi and Zhang, Xueyong and Espinosa-Marzal, Rosa M and Gewirth, Andrew A},
  journal={ACS Applied Energy Materials},
  volume={3},
  number={8},
  pages={8086--8094},
  year={2020},
  publisher={ACS Publications}
}

@article{li2022unconventional,
  title={Unconventional interfacial water structure of highly concentrated aqueous electrolytes at negative electrode polarizations},
  author={Li, Chao-Yu and Chen, Ming and Liu, Shuai and Lu, Xinyao and Meng, Jinhui and Yan, Jiawei and Abru{\~n}a, H{\'e}ctor D and Feng, Guang and Lian, Tianquan},
  journal={Nature Communications},
  volume={13},
  number={1},
  pages={5330},
  year={2022},
  publisher={Nature Publishing Group UK London}
}

@article{perez2017underscreening,
  title={Underscreening in concentrated electrolytes},
  author={Alpha A. Lee and Perez-Martinez, Carla S and Smith, Alexander M and Perkin, Susan},
  journal={Faraday discussions},
  volume={199},
  pages={239--259},
  year={2017},
  publisher={Royal Society of Chemistry}
}

@article{ichii2020solvation,
  title={Solvation structure on water-in-salt/mica interfaces and its molality dependence investigated by atomic force microscopy},
  author={Ichii, Takashi and Ichikawa, Satoshi and Yamada, Yuya and Murata, Makoto and Utsunomiya, Toru and Sugimura, Hiroyuki},
  journal={Japanese Journal of Applied Physics},
  volume={59},
  number={SN},
  pages={SN1003},
  year={2020},
  publisher={IOP Publishing}
}

@article{Orangi2024,
  title={Historical and prospective lithium-ion battery cost trajectories from a bottom-up production modeling perspective},
  author={Sina Orangi and Nelson Manjong and Daniel Perez Clos and Lorenzo Usai and Odne Stokke Burheim and Anders Hammer Stromman },
  journal={Journal of Energy Storage},
  volume= {76},
  pages={109800},
  year={2024},
}

@article{Wu2022,
  title={Significance of Antisolvents on Solvation Structures Enhancing Interfacial Chemistry in Localized High-Concentration Electrolytes},
  author={Yanzhou Wu and Aiping Wang and Qiao Hu and Hongmei Liang and Hong Xu and Li Wang and Xiangming He},
  journal={ACS Cent. Sci.},
  volume={8},
  pages={1290--1298},
  year={2022},
}

@article{Beltran2020LHCE,
  title={Localized High Concentration Electrolytes for High Voltage Lithium-Metal Batteries: Correlation between the Electrolyte Composition and Its Reductive/Oxidative Stability},
  author={Saul Perez Beltran and Xia Cao and Ji-Guang Zhang and Perla B. Balbuena},
  journal={Chem. Mater.},
  volume={32},
  pages={5973--5984},
  year={2020},
}

@misc{Goodenough2013ThePerspective,
    title = {{The Li-ion rechargeable battery: A perspective}},
    year = {2013},
    booktitle = {Journal of the American Chemical Society},
    author = {Goodenough, John B. and Park, Kyu Sung},
    number = {4},
    volume = {135},
    doi = {10.1021/ja3091438},
    issn = {00027863}
}

@article{Cresce2017,
  title={Solvation behavior of carbonate-based electrolytes in sodium ion batteries},
  author={Arthur V. Cresce and Selena M. Russell and Oleg Borodin and Joshua A. Allen and Marshall A. Schroeder and Michael Dai and Jing Peng and Mallory P. Gobet and Steven G. Greenbaum and Reginald E. Rogers and Kang Xu},
  journal={Phys. Chem. Chem. Phys.},
  volume={19},
  pages={574},
  year={2017},
}

@article{phelan2024role,
  title={Role of salt concentration in stabilizing charged Ni-rich cathode interfaces in Li-ion batteries},
  author={Phelan, Conor ME and Bj\"orklund, Erik and Singh, Jasper and Fraser, Michael and Didwal, Pravin N and Rees, Gregory J and Ruff, Zachary and Ferrer, Pilar and Grinter, David C and Grey, Clare P and others},
  journal={Chemistry of Materials},
  volume={36},
  number={7},
  pages={3334--3344},
  year={2024},
  publisher={ACS Publications}
}

@article{bjorklund2022cycle,
  title={Cycle-induced interfacial degradation and transition-metal cross-over in LiNi0. 8Mn0. 1Co0. 1O2-graphite cells},
  author={Bjorklund, Erik and Xu, Chao and Dose, Wesley M and Sole, Christopher G and Thakur, Pardeep K and Lee, Tien-Lin and De Volder, Michael FL and Grey, Clare P and Weatherup, Robert S},
  journal={Chemistry of Materials},
  volume={34},
  number={5},
  pages={2034--2048},
  year={2022},
  publisher={ACS Publications}
}

@article{rinkel2020electrolyte,
  title={Electrolyte oxidation pathways in lithium-ion batteries},
  author={Rinkel, Bernardine LD and Hall, David S and Temprano, Israel and Grey, Clare P},
  journal={Journal of the American Chemical Society},
  volume={142},
  number={35},
  pages={15058--15074},
  year={2020},
  publisher={ACS Publications}
}

@article{rinkel2022two,
  title={Two electrolyte decomposition pathways at nickel-rich cathode surfaces in lithium-ion batteries},
  author={Rinkel, Bernardine LD and Vivek, J Padmanabhan and Garcia-Araez, Nuria and Grey, Clare P},
  journal={Energy \& environmental science},
  volume={15},
  number={8},
  pages={3416--3438},
  year={2022},
  publisher={Royal Society of Chemistry}
}

@article{phelan2026linking,
  title={Linking solvation structure, activity, and electronic structure to electrochemical stability in lithium-ion battery electrolytes},
  author={Phelan, Conor and Bhandari, Arihant and Singh, Jasper and Fraser, Michael and Bj{\"o}rklund, Erik and Swallow, Jack and Bencok, Peter and Grey, Clare and Skylaris, Chris and Goodwin, Zachary and others},
  year={2026},
  journal={https://chemrxiv.org/doi/pdf/10.26434/chemrxiv-2025-k903g/v2}
}

\end{document}